\documentclass[fleqn, preprint,review,11pt]{elsarticle}
\usepackage[margin=1in]{geometry}

\usepackage{physics}
\usepackage{amsmath}
\usepackage{amssymb}
\usepackage{esint}
\usepackage{siunitx}
\usepackage{multicol}
\usepackage{pdflscape}
\usepackage{multirow}
\usepackage{multicol}
\usepackage{nicefrac}
\usepackage{booktabs}
\usepackage{float}
\usepackage{svg}
\usepackage{epstopdf}
\usepackage{xspace}
\usepackage[explicit]{titlesec}
\usepackage[normalem]{ulem}
\usepackage{cleveref}
\usepackage{xcolor}
\usepackage{verbatim}
\usepackage{mathrsfs}
\usepackage{hhline}
\usepackage{longtable}
\usepackage{tabu}
\usepackage{booktabs}
\usepackage{bm}
\usepackage{soul}
\setstcolor{red}
\usepackage{adjustbox}
\newlength{\fullpagewidth}
\usepackage{longtable}

\usepackage{lineno}
\usepackage{caption}
\usepackage[position=top]{subfig}
\makeatletter
\let\LN@align\align
\let\LN@endalign\endalign
\renewcommand{\align}{\linenomath\LN@align}
\renewcommand{\endalign}{\LN@endalign\endlinenomath}
\let\LN@gather\gather
\let\LN@endgather\endgather
\renewcommand{\gather}{\linenomath\LN@gather}
\renewcommand{\endgather}{\LN@endgather\endlinenomath}
\makeatother

\newcommand{\ensmean}[1]{\left\langle #1 \right\rangle}

\let\oldfrac\frac
\renewcommand{\frac}[2]{%
  \mathchoice
    {\oldfrac{#1}{#2}}
    {#1/#2}
    {#1/#2}
    {#1/#2}
}

\newcommand{\ddfrac}[2]{\frac{\displaystyle #1}{\displaystyle #2}}

\def\K{\mathbf{K}}

\def\n{\mathbf{n}}
\def\R{\mathbf{R}}

\def\U{\mathbf{U}}

\def\V{\mathbf{V}}

\def\x{\mathbf{x}}
\def\y{\mathbf{y}}

\def\chiv{\bm{\chi}}
\def\xiv{\bm{\xi}}

\def\I{\mathbf{I}}

\def\derive#1#2{\frac{\partial #1}{\partial #2}}

\def\neweq#1#2{\begin{equation} \label{#1}\begin{aligned} #2 \end{aligned}\end{equation}}
\def\neweqgat#1#2{\begin{equation} \label{#1}\begin{gathered} #2 \end{gathered}\end{equation}}

\def\xloop#1{\ifx\relax#1\else[#1]\expandafter\xloop\fi}

\def\lookfori#1{\ifx i#1 \expandafter\lookfors \else \expandafter\lookfori\fi}
\def\lookfors#1{\ifx s#1 \expandafter\lookforunderscore \else \expandafter\lookfori\fi}
\def\lookforunderscore#1{\ifx \_#1 \expandafter\savefirstletter \else \expandafter\lookfori\fi}
\def\savefirstletter#1{\def\letterone{#1}\expandafter\stopatnext}
\def\stopatnext#1{\ifx \_#1 \break\fi \savesecondletter#1}
\def\savesecondletter#1{\def\lettertwo{#1}\expandafter\stopatnext}

\DeclareMathOperator*{\argmax}{arg\,max}
\DeclareMathOperator*{\argmin}{arg\,min}

\newcommand\strev{\bgroup\markoverwith{\textcolor{red}{\rule[.5ex]{2pt}{0.4pt}}}\ULon}

\newcommand{\symbolica}{\textsf{Symbolica}\xspace}

\newcommand{\fenics}{\textsf{FEniCS}\xspace}

\journal{}

\begin{document}
\begin{frontmatter}
              
\title{Non-intrusive Auto-detecting and Adaptive Hybrid Scheme for Multiscale Heat Transfer: Thermal Runaway in a Battery Pack} 

\author[1,2,3]{Yinuo Noah Yao}
\cortext[cor1]{Corresponding author}
\author[1]{Ilenia Battiato\corref{cor1}}

\address[1]{Department of Energy Science and Engineering, Stanford University, Stanford, California, USA}
\address[2]{now at Department of Civil and Environmental Engineering, Texas A\& M University, College Station, Texas, USA}
\address[3]{Department of Mechanical Engineering, Texas A\& M University, College Station, Texas, USA}

\begin{abstract}
Accurately capturing and simulating multiscale systems is a formidable challenge, as both spatial and temporal scales can span many orders of magnitude. Rigorous upscaling methods not only ensure efficient computation, but also maintains errors within a priori prescribed limits. This provides a balance between computational costs and accuracy. However, the most significant difficulties arise when the conditions under which upscaled models can be applied cease to hold. To address this, we develop an automatic-detecting and adaptive, nonintrusive two-sided hybrid method for multiscale heat transfer and apply it to thermal runaway in a battery pack. To allow adaptive hybrid simulations, two kernels are developed to dynamically map the values between the fine-scale and the upscaled subdomains in a single simulation. The accuracy of the developed hybrid method is demonstrated through conducting a series of thermal runaway test cases in a battery pack. Our results show that the maximum spatial errors consistently remain below the threshold bounded by upscaling errors.
\end{abstract}


\end{frontmatter}

\begin{footnotesize}
\begin{longtable}{|p{0.15\fullpagewidth} p{0.35\fullpagewidth} p{0.15\fullpagewidth} p{0.35\fullpagewidth}|}
\hline
\textbf{Nomenclature} &  &  & \\ \hline
$\hat{\Omega}_\epsilon$ & Fine-scale representation of a two-dimensional battery pack & $\hat{\mathcal{B}}_\epsilon^{\left(c\right)}$, $\hat{\mathcal{B}}_\epsilon^{\left(p\right)}$, $\hat{\mathcal{B}}_\epsilon^{\left(w\right)}$ & Domains for the battery cell, packing material, and cooling water pipe in $\hat{\Omega}_\epsilon$ \\ 
$\hat{\Gamma}_\epsilon^{\left(pc\right)}$, $\hat{\Gamma}_\epsilon^{\left(pw\right)}$ & Interfaces for packing-cell and packing-pipe in $\hat{\Omega}_\epsilon$ & $\epsilon$ & Separation of scale parameter \\
$N^{\left(c\right)}_{x}$, $N^{\left(c\right)}_{y}$ & Number of unit cells in the $x$ and $y$ directions of the battery pack & $\hat{\mathcal{L}}_{x}$, $\hat{\mathcal{L}}_{y}$ & Length and width of the battery pack  [\si{L}] \\ 
$\hat{\ell}$ & Length of the unit cell  [\si{L}] & $a$ & Aspect ratio of the unit cell [\si{-}]\\
$\hat{r}_\epsilon^{\left(c\right)}$, $\hat{r}_\epsilon^{\left(w\right)}$ & Radius of the battery cell and cooling water pipe [\si{L}] & $\hat{d}_\epsilon^{\left(cc\right)}$, $\hat{d}_\epsilon^{\left(1\right)}$, $\hat{d}_\epsilon^{\left(2\right)}$ & Distances between the battery cell, edges of the unit cell, and cooling pipe [\si{L}] \\
$\hat{T}_{\epsilon}^{\left(p\right)}$, $\hat{T}_{\epsilon}^{\left(c\right)}$ & Temperatures of the packing material and battery cell [\si{\Theta}] & $\n_{\epsilon}^{\left(p\right)}$, $\n_{\epsilon}^{\left(c\right)}$ & Normal vectors to the interfaces, pointing away from the packing material and battery cell domains \\
$\hat{U}^{\left(pc\right)}$  & Total heat transfer coefficient between the packing material and battery cells [\si{MT\tothe{-3}\Theta\tothe{-1}}] & $\hat{q}_{\epsilon}^{\left(pw\right)}$ & Power flux between the packing material and cooling water pipes [\si{MT\tothe{-3}}] \\
$\hat{\Pi}$ & Power flux source term for the battery cell [\si{ML\tothe{-1}T\tothe{-3}}] & $\hat{\rho}^{\left(p\right)}$, $\hat{\rho}^{\left(c\right)}$ & Densities of the packing material and battery cell [\si{ML\tothe{-3}}] \\
$\hat{k}^{\left(p\right)}$, $\hat{k}^{\left(c\right)}$ & Thermal conductivities of the packing material and battery cell [\si{MLT\tothe{-3}\Theta\tothe{-1}}] & $\hat{C}^{\left(p\right)}$, $\hat{C}^{\left(c\right)}$ & Heat capacities of the packing material and battery cell [\si{L\tothe{2}T\tothe{-2}\Theta\tothe{-1}}] \\
$\hat{\Pi}_{\text{base}}$, $\hat{\Pi}_{\text{burn}}$ & \textit{Base} and \textit{burn} power flux values [\si{ML\tothe{-1}T\tothe{-3}}] & $\hat{T}_{ref}$ & Reference temperature [\si{\Theta}]\\
$\hat{T}_a$ & Temperature intervals within which the power flux source term $\hat{\Pi} = \hat{\Pi}_{\text{base}}$ [\si{\Theta}] & $\hat{T}_b$  & Temperature intervals within which the power flux source term $\hat{\Pi} = \hat{\Pi}_{\text{burn}}$ [\si{\Theta}] \\
$\hat{T}_{s1}$ & Temperature range over which $\hat{\Pi}$ transitions from $\hat{\Pi}_{\text{base}}$ to $\hat{\Pi}_{\text{burn}}$ [\si{\Theta}] & $\hat{T}_{s2}$ & Temperature range over which $\hat{\Pi}$ transitions from $\hat{\Pi}_{\text{burn}}$ to zero [\si{\Theta}] \\
$\epsilon_{s1}$, $\epsilon_{s2}$ & Dimensionless parameters governing the smoothness of the error functions [\si{-}] & $\hat{\mathcal{K}}^{\left(p\right)}$, $\hat{\mathcal{K}}^{\left(c\right)}$ & Reference scales of thermal conductivity for the packing material and battery cell [\si{MLT\tothe{-3}\Theta\tothe{-1}}] \\
$\hat{Q}^{\left(pw\right)}$ & Reference scale of the flux term at the $\hat{\Gamma}_{\epsilon}^{\left(pw\right)}$ boundary [\si{MT\tothe{-3}}] & $T^{\left(i\right)}_{HC}$ & Average temperature in the hybrid simulations \\
$\text{Bi}^{\left(p\right)}$, $\text{Bi}^{\left(c\right)}$ & Biot numbers for the packing material and battery cells (Dimensionless number) & $\mathcal{Q}$ & Ratio of conductance at the packing material-cooling pipe interface to the packing material conductance (Dimensionless number) \\
$\varrho$ & Product of density and heat capacity ratios (Dimensionless number) & $\varsigma$ & Thermal conductivity ratio (Dimensionless number) \\
$\mathcal{R}$ & Ratio of heat generation to the packing material conductance (Dimensionless number) & ${T}_{\epsilon}^{\left(p\right)}$, ${T}_{\epsilon}^{\left(c\right)}$ & Dimensionless temperatures of the packing material and battery cells [\si{-}] \\
$\Pi$ & Dimensionless power flux source term for the battery cells [\si{-}] & ${k}^{\left(p\right)}$, ${k}^{\left(c\right)}$ & Dimensionless thermal conductivities of the packing material and battery cells [\si{-}] \\
$\phi^{\left(p\right)}$, $\phi^{\left(c\right)}$ & Volume fractions of the packing material and battery cells in the unit cell [\si{-}] & $\langle T^{\left(p\right)} \rangle_{Y}$, $\langle T^{\left(c\right)} \rangle_{Y}$ & Average temperatures of the packing material and battery cells [\si{-}] \\
$\overline{\Pi}$ & Homogenized power flux source term [\si{-}] & ${\Pi}_{\text{base}}$ & Dimensionless \textit{base} power flux values [\si{-}] \\
$\U^{\left(p\right)}$, $\U^{\left(c\right)}$ & Effective velocities of the packing material and battery cells [\si{-}] & $\V^{\left(p\right)}$, $\V^{\left(c\right)}$ & Effective parameters corresponding to emergent terms for the packing material and battery cells [\si{-}] \\
$\K^{\left(p\right)}$, $\K^{\left(c\right)}$ & Effective thermal conductivities [\si{-}] & $R_{\bm{\cdot}}^{\left(p\right)}$, $R_{\bm{\cdot}}^{\left(c\right)}$, $\R_{5}^{\left(p\right)}$ & Effective reaction rates with $\cdot=\{1,2,3,4\}$ [\si{-}] \\
$\chi^{\left(p\right)\left[1\right]}$, $\chi^{\left(p\right)\left[2\right]}$, $\chiv^{\left(p\right)\left[3\right]}$ & Closure variables for the packing material & $\chi^{\left(c\right)\left[1\right]}$, $\chiv^{\left(c\right)\left[2\right]}$ & Closure variables for the battery cells \\
$\Omega_{\text{fine}}$, ${\Omega}_{\text{up}}$ & Fine-scale and upscaled subdomains in a two-dimensional battery pack & ${\Omega}_{\text{break}}$ & Breakdown region of the battery pack \\
$\mathcal{W}_{\epsilon}^{\left(p\right)}$, $\mathcal{W}_{\epsilon}^{\left(c\right)}$ & Moving average windows of the packing material and battery cells & $\Omega^{\text{old}}$, $\Omega^{\text{new}}$ & Battery pack domain before and after the expansion or reduction of the breakdown region \\
$\Omega_{\text{fine}}^{\text{old}}$, $\Omega_{\text{up}}^{\text{old}}$ & Fine-scale and upscaled subdomains before the expansion or reduction of the breakdown region & $\Omega_{\text{fine}}^{\text{new}}$, $\Omega_{\text{up}}^{\text{new}}$ & Fine-scale and upscaled subdomains after the expansion or reduction of the breakdown region \\
$\Theta^{\left(p\right)}_{\epsilon}$, $\Theta^{\left(c\right)}_{\epsilon}$ & Temperatures of the packing material and battery cells & $\langle \Theta^{\left(i\right)} \rangle_{Y}$, $\langle \Theta^{\left(i\right)} \rangle_{Y}$ & Average temperatures of the packing material and battery cells \\
$\mathcal{V}_{\text{up}}$, $\mathcal{V}_{\text{fine}}$ & Common upscaled and fine-scale subdomains present in both $\Omega^{\text{old}}$ and $\Omega^{\text{new}}$ & $\mathcal{V}_{\text{up} \rightarrow \text{fine}}$ & Upscaled subdomain in $\Omega^{\text{old}}$ that is switched to fine-scale in $\Omega^{\text{new}}$ \\
$\mathcal{V}_{\text{fine} \rightarrow \text{up}}$ & Fine-scale subdomain in $\Omega^{\text{old}}$ that is switched to upscaled in $\Omega^{\text{new}}$ & $\Pi_{\text{NB}}$, $\Pi_{\text{FB}}$ & Dimensionless fine-scale power flux source terms for unburned and burning battery cells \\
$\gamma$ & Arbitrary constant for approximating a stepwise function & $\overline{\Pi}_{\text{NB}}$, $\overline{\Pi}_{\text{FB}}$ & Dimensionless upscaled power flux source terms for unburned and burning battery cells \\
\hline
\end{longtable}
\end{footnotesize}

\section{Introduction}
As societies attempt to move toward a sustainable energy transition, the importance of energy storage both for transportation and long term storage applications cannot be overstated. A continuing challenge in battery energy storage  is related to the hazard of explosion due to thermal runaway as a result of mechanical, thermal, and electric abuse~\cite{Feng2018-ix, Wang2012-cd, Simunovic2020-qh, Noii2024-do}. Understanding and predicting heat transfer and thermal runaway within a battery pack at the relevant scales becomes critical to optimize battery pack design and management. Yet, model development for accurate prediction of heat transfer in battery packs, ranging from the subelectrode to the battery pack scale, remains a formidable task due to the complex interactions between processes across scales, ranging from the $\mu$m to the m scale and beyond. The challenge of accurate modeling of multiscale multiphysics systems is not unique to batteries, and, in fact, many advancements both theoretical and algorithmic (numerical and symbolic) have been made since the importance of multiscale effects was first recognized in solid mechanics in the 70ies, and later permeated to a variety of  fields ranging from transport and reactions in geologic to biologic porous media, just to mention a few applications \cite{Battiato2019-xk}.

Differently from other fields which have ripen the benefits of more sophisticated modeling approaches for multiscale systems, the use of advanced computer-aided design and multiscale models to better understand, predict and optimize battery behavior has been generally overshadowed by new materials discovery and the needs of the fast expanding EV transportation market, where equivalent circuit and P2D  models still dominate. Yet, increasingly recognized performance limitations of such models  
 has led to a renewed interest in more sophisticated approaches to capture multiscale physics effects more accurately \cite{Aruchalam-2015-temperature,Pietrzyk2023-ou}. 

Coarse-graining techniques include a suite of mathematical theories that allow to rigrously derive  continuum-scale models and multiscale formulations from fine-grained models (for a review of different methods, see e.g. \cite{Battiato2019-xk}). Among others, homogenization theory by multiple scale expansions has been sucessfully used to derive upscaled models  from fine-scale governing equations with \emph{a priori} error guarantees on accuracy~\cite{Langlo1994-wq, Vasilyeva2019-wh, Das2005-gm, Frippiat2008-cl, Battiato2019-xk, Battiato2011-ad, Pietrzyk2023-ou, Pietrzyk2021-lu, Pietrzyk2023-am, Pietrzyk2023-jy}. The applicability regimes of such upscaled equations, within which their accuracy can be rigorously guaranteed,  are dictated by a group of dimensionless numbers, which describe the intrinsic dynamics of the system. The main advantages of the homogenization/coarse-graining approach lie in its ability to (1) significantly reduce computational costs compared to solving fine-scale equations over the same computational domain, and (2) offer \emph{a priori} guaranteed error bounds within defined applicability regimes. However, the complexity involved in deriving upscaled equations has historically restricted the accessibility of this and other rigorous techniques to a broader community. 

Recent advancements in symbolic computing and automated symbolic deduction have contributed to remove some of the human-centered barriers in the derivation and adoption of rigorously derived multiscale models for systems of realistic complexity \cite{Pietrzyk2021-lu,Pietrzyk2023-am, Pietrzyk2023-jy, Pietrzyk2023-ou}, for which such derivations would be too complex, long and error-prone.  Pietrzyk \emph{et al.}~\cite{Pietrzyk2021-lu} have addressed this issue by developing an automated symbolic deduction pipeline, \symbolica, which automatically performs homogenization by multiple-scale expansion to derive upscaled equations from fine-scale equations with minimal human interaction, therefore, enhancing the accessibility of this mathematical technique to a broader community and enabling its extension to more complex applications. Furthermore, Pietrzyk \emph{et al.}~\cite{Pietrzyk2021-lu, Pietrzyk2023-am, Pietrzyk2023-jy, Pietrzyk2023-ou} have demonstrated the versality of \symbolica in various types of problems, ranging from reactive transport in porous media to heat transfer in battery packs. Nonetheless, one persistent challenge associated with the homogenization approach is the existence of breakdown regions that invalidate the upscaled equations, if some of the applicability conditions are violated \cite{Battiato2011-zo, Boso2013-ag, Pietrzyk2023-am}.

To overcome such challenges, hybrid methods have been developed~\cite{Battiato2011-zo, Yousefzadeh2017-yc, Yao2023-zi,Wang2024-on} in which fine-scale and upscaled equations are coupled in the same computational domain, with the fine-scale equations computed for the regions that violate the applicability regimes while upscaled equations are solved elsewhere. Hybrid methods, also  referred to as algorithm refinement techniques, can be generally classified into two broad categories, based on the type of coupling: intrusive and non-intrusive. Intrusive methods usually consist of ``overlapped" or ``handshake" regions where both fine-scale and upscaled models are concurrently solved~\cite[e.g.][]{Battiato2011-zo, Pettersson2013-ou}. This enables a direct coupling between the fine-scale and the upscaled subdomains at the expense of higher computational costs and implementation difficulty in legacy codes. In contrast, non-intrusive methods, based on domain decomposition approaches, employ boundary conditions to couple two subdomains together~\cite[e.g.][]{Yousefzadeh2017-yc, Yao2023-zi}. A literature review of multiscale formulations focused on application to flow and reactive transport in media with heterogeneous inclusions (e.g. porous media) can be found in \cite{Scheibe-2015-MAP}.

Differently from most current adaptive approaches focused on grid-refinement strategies or on multiphysics problems (i.e. where different physics is solved in different subdomains \cite[e.g.][]{Jenny2005-vm, Jenny2006-yu, Hajibeygi2011-am, Lee2009-kx}) where the criteria for adaptation are usually problem-specific and do not consider the applicability of the upscaled governing equations, here we focus on building adaptive hybrid schemes where the same physics is solved at different scales (i.e. a fine and a coarse-scale), because of the local breakdown of the approximations underlying the formulation of continuum scale models. Fixed hybrid methods \cite{Yao2023-zi}, where the size of the fine-scale subdomain is predetermined and remains constant throughout the simulation, have been demonstrated to maintain predictive accuracy while achieving speedups ranging from 3 to 15 times compared to pure fine-scale models for relatively small macroscopic domains, with the expected speedup increasing with the size of the simulation domain. Differently from  fixed hybrids in which the size of the fine-scale subdomain is selected as large as possible to ensure that the violation of applicability conditions is accommodated throughout the duration of the simulation, the adaptive hybrid allows the size of the fine-scale domain to always be as small as strictly needed. Since the computational cost of hybrid models (adaptive or fixed) is controlled by the size of the pore-scale domain, the adaptive hybrid will provide a larger speedup than a fixed hybrid, depending on the ratio between the size of the pore-scale domains used in either method. In addition, the adaptivity criteria and hybridization strategy are tightly integrated and based on theoretical upscaling methods to ensure that the entire framework (analytical upscaling, adaptivity criteria and hybridization conditions) is self-consistent and to guarantee that the adaptive hybrid simulation error is always \emph{a priori} bounded by the upscaling error prescribed by the theory, while minimizing the computational cost required to solve the local breakdown. Because the adaptive hybrid maintains the same accuracy of the fixed hybrid, but at a fraction of the computational cost, the reduction in error would be similar to that reported by \cite{Yao2023-zi} for similar setups. We apply these new strategies to model thermal runaway in spatially heterogenous battery packs, an application for which many of the numerical, symbolic and theoretical advancements mentioned above remain still largely unknown to the community \cite{Wang2024-dw}.


Hence, this study has two primary objectives. First, we focus on advancing algorithmic capabilities in the context of hybridization strategies based on homogenization theory, and develop an automatic-detecting and adaptive two-sided hybrid formulation that allows the size of the fine-scale subdomains to vary over time, while ensuring that the coupling error introduced   remains bounded at all times by the uscaling error. This aspect ensures that 
computational efficiency and predictive accuracy can be concurrently preserved. Second, we apply this novel algorithmic strategy to model thermal runaway in battery packs. Specifically, we couple the continuum scale PDEs describing thermal runaway at the pack level, derived through the automated symbolic deduction engine, \symbolica \cite{Pietrzyk2023-ou}, with their fine-scale (cell-scale) counterpart, and demonstrate the relevance and robustness of these strategies for battery modeling applications. The computational domain consists of battery cells, embedded in a packing domain, with spatially heterogenous properties to mimic system level variability due to manufacturing defects, differential aging between cells, etc.

The manuscript is organized as follows. \Cref{sec:fine-govern-eqs,sec:upscaled-govern-eqs} provide an overview of the fine-scale and upscaled governing equations. In \Cref{sec:TS_fixed_formulation}, we present the development of a two-sided hybridization approach with fixed coupling boundaries. \Cref{sec:ad-formulation,sec:kernels} detail the formulation of mapping kernels and techniques for automatic detection of the existence or expansion/reduction of the fine-scale subdomain in response to the varying breakdown region. In \Cref{sec:res_discuss}, we evaluate the accuracy of the proposed hybridization method with a simulated thermal runaway problem in a battery pack.

\section{Problem Setup and Model Formulation}

\begin{figure}
    \centering
    \includegraphics[width=\textwidth]{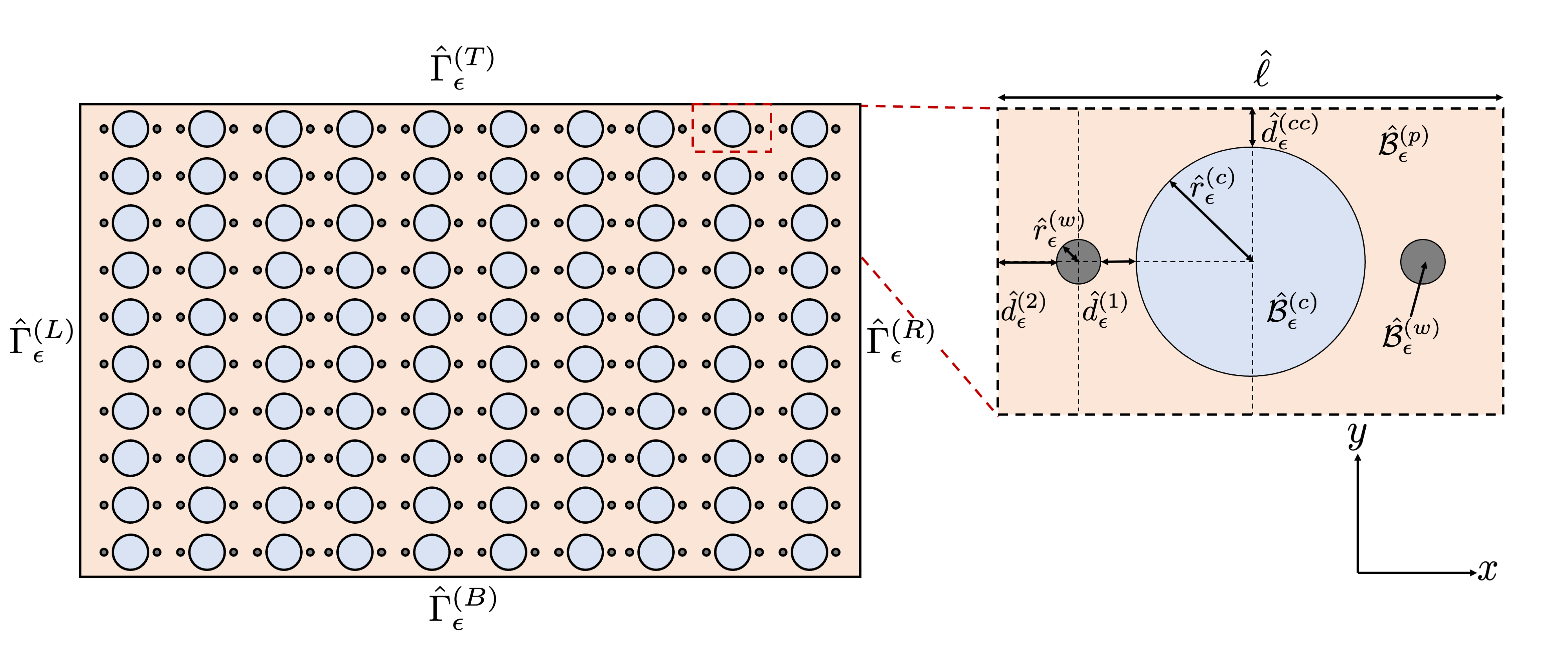}
    \caption{{Schematic diagram of a conceptual fine-scale battery pack.}}
    \label{fig:finescale-domain}
\end{figure}

{We consider a two-dimensional battery pack $\hat{\Omega}_\epsilon$ that comprises three distinct regions: (1) battery cells $\hat{\mathcal{B}}_\epsilon^{\left(c\right)} \subset \hat{\Omega}_\epsilon$, (2) packing material $\hat{\mathcal{B}}_\epsilon^{\left(p\right)} \subset \hat{\Omega}_\epsilon$, and (3) cooling water pipes $\hat{\mathcal{B}}_\epsilon^{\left(w\right)} \subset \hat{\Omega}_\epsilon$. $\hat{\Gamma}_\epsilon^{\left(pc\right)}$ and $\hat{\Gamma}_\epsilon^{\left(pw\right)}$ are the packing-cell and packing-pipe interfaces, respectively (\Cref{fig:finescale-domain}). Consistent with the notation in~\cite{Pietrzyk2023-ou, Yao2023-zi}, dimensional and fine-scale variables are hatted quantities and possess a subscript ``$\epsilon$", respectively.

\Cref{fig:finescale-domain} illustrates the unit-cell geometry that defines the geometry of the battery pack, with $N^{\left(c\right)}_{x}$ and $N^{\left(c\right)}_{y}$ representing the number of unit cells along the $x$ and $y$ directions, respectively. The length of the unit cell, $\hat{\ell}$, and the aspect ratio, $a$, are given by}

\begin{subequations}
\begin{align}
\hat{\ell} &= 2\left( \hat{d}_\epsilon^{\left(1\right)} + \hat{d}_\epsilon^{\left(2\right)} + \hat{r}_\epsilon^{\left(c\right)} + \hat{r}_\epsilon^{\left(w\right)} \right), \\
a &= \ddfrac{2\left(\hat{d}_\epsilon^{\left(cc\right)} + \hat{r}_\epsilon^{\left(c\right)}\right)}{\hat{\ell}} \label{eq:aspect_ratio},
\end{align}
\end{subequations}

\noindent where $\hat{r}_\epsilon^{\left(c\right)}$ and $\hat{r}_\epsilon^{\left(w\right)}$ are the radius of battery cell and cooling water pipe, respectively, and $\hat{d}_\epsilon^{\left(cc\right)}$, $\hat{d}_\epsilon^{\left(1\right)}$ and $\hat{d}_\epsilon^{\left(2\right)}$ are the distances between the battery cell, edges of the unit cell and the cooling pipe. The length and the width  of the whole battery pack, $\hat{\mathcal{L}}_{x}$ and $\hat{\mathcal{L}}_{y}$, respectively, are defined as: 
\begin{subequations}
\begin{align}
\hat{\mathcal{L}}_{x} &= N_{x}^{\left(c\right)}\hat{\ell},\\
\hat{\mathcal{L}}_{y} &= N_{y}^{\left(c\right)}\hat{\ell}a. \label{eq:eps}
\end{align}
\end{subequations}

Also,  the length scale ratio between the unit cell and the  battery pack is a fundamental parameter in multiscale systems, is referred to as separation of scale parameter $\epsilon$ and is defined as
\begin{align}
\epsilon = \frac{\ell}{\max \left(\hat{\mathcal{L}}_{x}, \hat{\mathcal{L}}_{y} \right)}.
\end{align}

\subsection{Dimensional and dimensionless fine-scale governing equations}
\label{sec:fine-govern-eqs}

\begin{figure}
    \centering
    \includegraphics[width=0.7\textwidth]{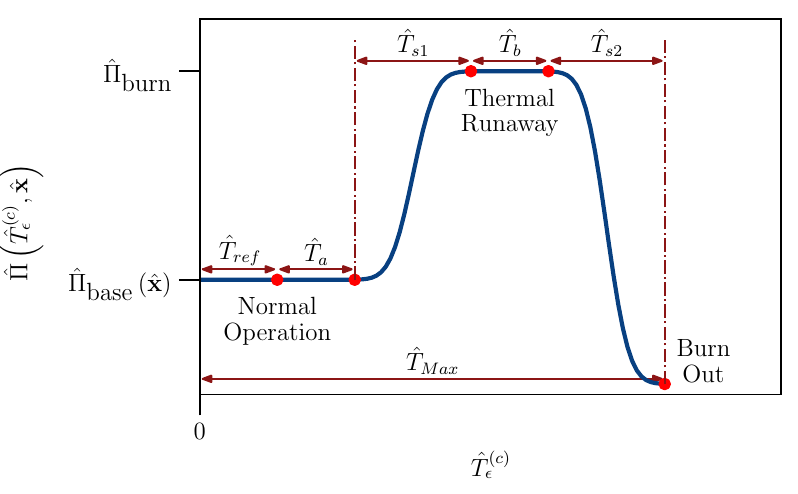}
    \caption{The profile for $\hat{\Pi}(\hat{T}_{\epsilon}^{\left(c\right)}, \hat{\mathbf{x}})$ as a function of the cell temperature $(\hat{T}_{\epsilon}^{\left(c\right)}$ (figure adapted from \cite{Pietrzyk2023-ou}) The lengths of the profile sections are shown to sum to $\hat{T}_{Max}$. Burnout is defined as occurring when $\hat{T}_{\epsilon}^{\left(c\right)} > \hat{T}_{\text{Max}}$ and $\hat{\Pi}(\hat{T}_{\epsilon}^{\left(c\right)}, \hat{\mathbf{x}}) = 0$}.
    \label{fig:Pi_Burn}
\end{figure}

In order to describe thermal conduction between cells and packing materials, and heat generation within cells as a function of temperature under normal and abuse (thermal runaway) conditions, we employ the model presented in Pietrzyk \emph{et al.} \cite{Pietrzyk2023-ou}, which we report here for completeness of the presentation. The packing material and battery cell temperatures at the fine/cell-scale are governed by two conduction equations
\begin{subequations}
\label{eq:Heat_Transfer_Eq_All}
\begin{align}
&\derive{\left(\hat{\rho}^{\left(p\right)}\hat{C}^{\left(p\right)}\hat{T}_{\epsilon}^{\left(p\right)}\right)}{\hat{t}} = \hat{\nabla}\bm{\cdot}\left(\hat{k}^{\left(p\right)}\hat{\nabla}\hat{T}_{\epsilon}^{\left(p\right)}\right), \quad \hat{\mathbf{x}} \in \hat{\mathcal{B}}_{\epsilon}^{\left(p\right)}, \label{eq:Heat_Transfer_Eq_Packing} \\
&\derive{\left(\hat{\rho}^{\left(c\right)}\hat{C}^{\left(c\right)}\hat{T}_{\epsilon}^{\left(c\right)}\right)}{\hat{t}} = \hat{\nabla}\bm{\cdot}\left(\hat{k}^{\left(c\right)}\hat{\nabla}\hat{T}_{\epsilon}^{\left(c\right)}\right) + {\hat{\Pi}}, \quad \hat{\mathbf{x}} \in \hat{\mathcal{B}}_{\epsilon}^{\left(c\right)}, \label{eq:Heat_Transfer_Eq_Cell} 
\end{align}
\noindent subject to the following boundary conditions
\begin{align}
 &-\n_{\epsilon}^{\left(p\right)} \bm{\cdot} \hat{k}^{\left(p\right)}\hat{\nabla}\hat{T}_{\epsilon}^{\left(p\right)} = \hat{U}^{\left(pc\right)}\left(\hat{T}_{\epsilon}^{\left(p\right)} - \hat{T}_{\epsilon}^{\left(c\right)}\right), \quad \hat{\mathbf{x}} \in \hat{\Gamma}_{\epsilon}^{\left(pc\right)}, \label{eq:BC_pc_packing}\\
&-\n_{\epsilon}^{\left(p\right)} \bm{\cdot} \hat{k}^{\left(p\right)}\hat{\nabla}\hat{T}_{\epsilon}^{\left(p\right)} = {\hat{q}_{\epsilon}^{\left(pw\right)}}, \quad  \hat{\mathbf{x}} \in \hat{\Gamma}_{\epsilon}^{\left(pw\right)}, \label{eq:BC_pw}\\
&-\n_{\epsilon}^{\left(c\right)} \bm{\cdot} \hat{k}^{\left(c\right)}\hat{\nabla}\hat{T}_{\epsilon}^{\left(c\right)} = \hat{U}^{\left(pc\right)}\left(\hat{T}_{\epsilon}^{\left(c\right)} - \hat{T}_{\epsilon}^{\left(p\right)}\right), \quad  \hat{\mathbf{x}} \in \hat{\Gamma}_{\epsilon}^{\left(pc\right)}, \label{eq:BC_pc_cell}  
\end{align}
\end{subequations}

\noindent where $\hat{T}_{\epsilon}^{\left(i\right)}  \equiv \hat{T}_{\epsilon}^{\left(i\right)}\left(\hat{t}, \hat{\mathbf{x}}\right) [\si{\Theta}]$ is the temperature at time $\hat{t}$ and location $\hat{\mathbf{x}} \in \hat{\mathcal{B}}_{\epsilon}^{\left(i\right)}$, and  $i=p$ or $c$ refers to packing material or battery cell, respectively; $\n_{\epsilon}^{\left(i\right)} \equiv \n_{\epsilon}^{\left(i\right)}\left(\hat{\mathbf{x}}\right)$ is the normal vector to the interfaces pointing away from the $i$ domain; $\hat{U}^{\left(pc\right)}$ [\si{MT\tothe{-3}\Theta\tothe{-1}}] is the total heat transfer coefficient between the packing material and battery cells;  $\hat{q}_{\epsilon}^{\left(pw\right)}\left(\hat{t}, \hat{\mathbf{x}}\right)$ [\si{MT\tothe{-3}}]  is a power flux between the packing material and the cooling water pipes; $\hat{\Pi}(\hat{t}, \hat{\mathbf{x}})$ [\si{ML\tothe{-1}T\tothe{-3}}] is a power flux source term, and $\hat{\rho}^{\left(i\right)}$ [\si{ML\tothe{-3}}], $\hat{k}^{\left(i\right)}$ [\si{MLT\tothe{-3}\Theta\tothe{-1}}] and $\hat{C}^{\left(i\right)}$ [\si{L\tothe{2}T\tothe{-2}\Theta\tothe{-1}}] are density, thermal conductivity, and heat capacity, respectively. {The heat generation within the battery cells is modeled with the power flux source term $\hat{\Pi}(\hat{t}, \hat{\mathbf{x}})$ (\Cref{fig:Pi_Burn}), and is given by}

\begin{subequations}
\label{eq:PI}
\begin{align}
\hat{\Pi}\left(\hat{T}_{\epsilon}^{\left(c\right)}, \hat{\mathbf{x}}\right) &= \hat{\Pi}_{\text{base}}\left(\hat{\mathbf{x}}\right) + \frac{1}{2}\left\{\text{Erf}\left[C_1\left(2\frac{\hat{T}_a - \hat{T}_{\epsilon}^{\left(c\right)} + \hat{T}_{ref}}{\hat{T}_{s1}} + 1\right)\right] + 1\right\}\left(\hat{\Pi}_{\text{burn}} - \hat{\Pi}_{\text{base}}\left(\hat{\mathbf{x}}\right)\right) \nonumber \\   
&- \frac{1}{2}\left\{\text{Erf}\left[C_2\left(2\frac{\hat{T}_{max} - \hat{T}_{\epsilon}^{\left(c\right)} + \hat{T}_{ref}}{\hat{T}_{s2}} - 1\right)\right] + 1\right\}\hat{\Pi}_{\text{burn}}, \label{eq:new_pi_term1}
\end{align}
\noindent where
\begin{align}
&\hat{T}_{max} = \hat{T}_{Max} - \hat{T}_{ref}, \label{eq:new_pi_term2}\\ 
&\hat{T}_{Max} = \hat{T}_{ref} + \hat{T}_{a} + \hat{T}_{s1} + \hat{T}_{b} + \hat{T}_{s2}, \label{eq:new_pi_term2_1}\\
&{C_1 = \text{Erf}^{-1}\left(2\epsilon_{s1} - 1\right),} \label{eq:new_pi_term3}\\
&{C_2 = \text{Erf}^{-1}\left(2\epsilon_{s2} - 1\right)}. \label{eq:new_pi_term3_2}  
\end{align}
\end{subequations}

In \Cref{eq:PI}, $\hat{\Pi}_{\text{base}}(\hat{\mathbf{x}})$ and $\hat{\Pi}_{\text{burn}}$ are the \textit{base} and \textit{burn} power flux values, respectively, $\hat{T}_{ref}$~[\si{\Theta}] is the reference temperature,  $\hat{T}_a$[\unit{\Theta}] and $\hat{T}_b$[\unit{\Theta}] are defined as the temperature intervals within which the power flux source term $\hat{\Pi}(\hat{T}_{\epsilon}^{\left(c\right)}, \hat{\mathbf{x}})$ equals to $\hat{\Pi}_{\text{base}}(\hat{\mathbf{x}})$ and $\hat{\Pi}_{\text{burn}}$, respectively. $\hat{T}_{s1}$ and $\hat{T}_{s2}$ refer to the temperature ranges over which $\hat{\Pi}(\hat{T}_{\epsilon}^{\left(c\right)}, \hat{\mathbf{x}})$ transitions from $\hat{\Pi}_{\text{base}}(\hat{\mathbf{x}})$ to $\hat{\Pi}_{\text{burn}}$, and from $\hat{\Pi}_{\text{burn}}$ to zero, respectively.  $\epsilon_{s1} = 0.0005$ and $\epsilon_{s2} = 0.0005$ are the dimensionless parameters that govern the smoothness of the error functions.
The system of equations \Cref{eq:Heat_Transfer_Eq_All,eq:PI} can be cast in dimensionless form by employing the following reference scales \cite{Pietrzyk2023-ou}:

\begin{align}
\label{eq:Scales}
\hat{T}_{\epsilon}^{\left(p\right)} = \left( \hat{T}_{a} + \hat{T}_{s1} + \hat{T}_{b} + \hat{T}_{s2} \right)T_{\epsilon}^{\left(p\right)} + \hat{T}_{ref}, \quad \hat{T}_{\epsilon}^{\left(c\right)} = \left( \hat{T}_{a} + \hat{T}_{s1} + \hat{T}_{b} + \hat{T}_{s2} \right)T_{\epsilon}^{\left(c\right)} + \hat{T}_{ref}, \nonumber \\
\hat{k}^{\left(p\right)} = \hat{\mathcal{K}}^{\left(p\right)} k^{\left(p\right)}, \quad \hat{k}^{\left(c\right)} = \hat{\mathcal{K}}^{\left(c\right)} k^{\left(c\right)}, \quad \hat{\nabla} = \frac{1}{\max(\hat{\mathcal{L}}_{x}, \hat{\mathcal{L}}_{y})} \nabla, \quad \hat{t} = \frac{\hat{\rho}^{\left(p\right)}\hat{C}^{\left(p\right)}{\max(\hat{\mathcal{L}}_{x}, \hat{\mathcal{L}}_{y})}^2}{\hat{\mathcal{K}}^{\left(p\right)}} t, \\  
\hat{\mathbf{x}} = \max(\hat{\mathcal{L}}_{x}, \hat{\mathcal{L}}_{y}) \mathbf{x},  \quad  \hat{\Pi}\left(\hat{T}_{\epsilon}^{\left(c\right)}, \hat{\mathbf{x}}\right) = \hat{\Pi}_{\text{burn}} \Pi\left(T_{\epsilon}^{\left(c\right)}, \mathbf{x}\right), \quad \hat{q}_{\epsilon}^{\left(pw\right)}\left(\hat{t}, \hat{\mathbf{x}}\right) = \hat{Q}^{\left(pw\right)} q_{\epsilon}^{\left(pw\right)}\left(t, \mathbf{x}\right), \nonumber
\end{align}

\noindent where $\hat{\mathcal{K}}^{\left(i\right)}$ is the reference scale of thermal conductivity, and $\hat{Q}^{\left(pw\right)}$ is the reference scale of the flux term at the $\hat{\Gamma}_{\epsilon}^{\left(pw\right)}$ boundary. Using \Cref{eq:Scales} to nondimensionalize the fine-scale governing equations \Cref{eq:Heat_Transfer_Eq_All,eq:PI}, the dimensionless fine-scale governing equations can be written as~\cite{Pietrzyk2023-ou} 

\begin{subequations}
\begin{align}
&\derive{T_{\epsilon}^{\left(p\right)}}{t} = \nabla \bm{\cdot} \left(k^{\left(p\right)} \nabla T_{\epsilon}^{\left(p\right)}\right) \quad \text{for } \mathbf{x} \in \mathcal{B}_{\epsilon}^{\left(p\right)}, \label{eq:Heat_Transfer_Eq_Packing_dimless} \\
&\derive{T_{\epsilon}^{\left(c\right)}}{t} = \left(\varrho \varsigma\right) \nabla \bm{\cdot} \left(k^{\left(c\right)} \nabla T_{\epsilon}^{\left(c\right)}\right) + \left(\varrho  \mathcal{R}\right) {\Pi} \quad \text{for } \mathbf{x} \in \mathcal{B}_{\epsilon}^{\left(c\right)}, \label{eq:Heat_Transfer_Eq_Cell_dimless}
\end{align}
\label{eq:pore_goven_eqs}
\noindent subject to boundary conditions 
\begin{align}
&-\n_{\epsilon}^{\left(p\right)} \bm{\cdot} k^{\left(p\right)} \nabla T_{\epsilon}^{\left(p\right)} = \text{Bi}^{\left(p\right)}\left(T_{\epsilon}^{\left(p\right)} - T_{\epsilon}^{\left(c\right)}\right) \quad \text{for } \mathbf{x} \in \Gamma_{\epsilon}^{\left(pc\right)}, \label{eq:BC_pc_packing_dimless} \\
&-\n_{\epsilon}^{\left(p\right)} \bm{\cdot} k^{\left(p\right)}\nabla T_{\epsilon}^{\left(p\right)} = \mathcal{Q}  {q_{\epsilon}^{\left(pw\right)}} \quad \text{for } \mathbf{x} \in \Gamma_{\epsilon}^{\left(pw\right)}, \label{eq:BC_pw_dimless} \\
&-\n_{\epsilon}^{\left(c\right)} \bm{\cdot} k^{\left(c\right)} \nabla T_{\epsilon}^{\left(c\right)} = \text{Bi}^{\left(c\right)}\left(T_{\epsilon}^{\left(c\right)} - T_{\epsilon}^{\left(p\right)}\right) \quad \text{for } \mathbf{x} \in \Gamma_{\epsilon}^{\left(pc\right)}. \label{eq:BC_pc_cell_dimless} 
\end{align}
\end{subequations}
\noindent The dimensionless power flux source term $\Pi$ is defined as
\begin{subequations}
\begin{align}
\Pi(T_{\epsilon}^{\left(c\right)}, \mathbf{x}) &= \Pi_{\text{base}}\left(\mathbf{x}\right) + \frac{1}{2}\left\{\text{Erf}\left[A_1 T_{\epsilon}^{\left(c\right)} + B_1\right] + 1\right\}\left(1 - \Pi_{\text{base}}\left(\mathbf{x}\right)\right) \nonumber \\ 
& - \frac{1}{2}\left\{\text{Erf}\left[A_2 T_{\epsilon}^{\left(c\right)} + B_2\right] + 1\right\},\label{eq:new_pi_term_dimless}
\end{align}
\noindent where
\begin{gather}
A_1 = -2C_1\frac{\hat{T}_{max}}{\hat{T}_{s1}}, \quad B_1 = 2C_1\frac{\hat{T}_{a}}{\hat{T}_{s1}} + C_1, \quad A_2 = -2C_2\frac{\hat{T}_{max}}{\hat{T}_{s2}},  \nonumber \\ B_2 = 2C_2\frac{\hat{T}_{max}}{\hat{T}_{s2}} - C_2, \quad C_1 = \text{Erf}^{-1}\left(2\epsilon_{s1} - 1\right), \quad {C_2 = \text{Erf}^{-1}\left(2\epsilon_{s2} - 1\right)}.
\end{gather}
\end{subequations}
\noindent By nondimensionalizing the governing equations,  six dimensionless numbers are defined as
\neweqgat{eq:dimless_groups}{\text{Bi}^{\left(p\right)} = \frac{\hat{U}^{\left(pc\right)}\max(\hat{\mathcal{L}}_{x}, \hat{\mathcal{L}}_{y})}{\hat{\mathcal{K}}^{\left(p\right)}}, \quad \mathcal{Q} = \frac{\hat{Q}^{\left(pw\right)}\max(\hat{\mathcal{L}}_{x}, \hat{\mathcal{L}}_{y})}{\hat{T}_{max}\hat{\mathcal{K}}^{\left(p\right)}}, \quad \varrho = \frac{\hat{\rho}^{\left(p\right)}\hat{C}^{\left(p\right)}}{\hat{\rho}^{\left(c\right)}\hat{C}^{\left(c\right)}}, \\
\varsigma = \frac{\hat{\mathcal{K}}^{\left(c\right)}}{\hat{\mathcal{K}}^{\left(p\right)}},  \quad \text{Bi}^{\left(c\right)} = \frac{\text{Bi}^{\left(p\right)}}{\varsigma}, \quad \mathcal{R} = \frac{\hat{\Pi}_{\text{burn}}\max(\hat{\mathcal{L}}_{x}, \hat{\mathcal{L}}_{y})^2}{\hat{T}_{max}\hat{\mathcal{K}}^{\left(p\right)}}.}

\subsection{Dimensionless upscaled governing equations}
\label{sec:upscaled-govern-eqs}

Pietrzyk \emph{et al.} \cite{Pietrzyk2023-ou} automatically upscaled the fine-scale governing equations presented in \Cref{sec:fine-govern-eqs} by means of homogenization theory \cite{Battiato2019-xk} using \symbolica \cite{Pietrzyk2021-lu} to obtain the following upscaled governing equations for the average packing material and cell temperatures, $\langle T^{\left(p\right)} \rangle_{Y}\left(t, \mathbf{x}\right)$ and $\langle T^{\left(c\right)} \rangle_{Y}\left(t, \mathbf{x}\right)$, respectively,

\begin{subequations}\label{upscaled}
\label{eq:upscaled_goven_eqs}
\neweq{Nc_2_homo_eq_1}{\phi^{\left(p\right)}\derive{\langle T^{\left(p\right)} \rangle_{Y}}{t} + \U^{\left(p\right)} \bm{\cdot} \nabla_{\mathbf{x}} \langle T^{\left(p\right)} \rangle_{Y} - \V^{\left(p\right)} \bm{\cdot} \nabla_{\mathbf{x}} \langle T^{\left(c\right)} \rangle_{Y} - \nabla_{\mathbf{x}} \bm{\cdot} \left(\K^{\left(p\right)} \bm{\cdot} \nabla_{\mathbf{x}} \langle T^{\left(p\right)} \rangle_{Y}\right)                   \\
= -R_1^{\left(p\right)}\langle T^{\left(p\right)} \rangle_{Y} + R_2^{\left(p\right)}\langle T^{\left(c\right)} \rangle_{Y} - R_3^{\left(p\right)}q^{\left(pw\right)}\left(t, \mathbf{x}\right) + \R_4^{\left(p\right)} \bm{\cdot} \nabla_{\mathbf{x}} {q^{\left(pw\right)}},}

\neweq{Nc_2_homo_eq_2}{\phi^{\left(c\right)}\derive{\langle T^{\left(c\right)} \rangle_{Y}}{t} + \U^{\left(c\right)} \bm{\cdot} \nabla_{\mathbf{x}} \langle T^{\left(c\right)} \rangle_{Y} - \V^{\left(c\right)} \bm{\cdot} \nabla_{\mathbf{x}} \langle T^{\left(p\right)} \rangle_{Y} - \nabla_{\mathbf{x}} \bm{\cdot} \left(\K^{\left(c\right)} \bm{\cdot} \nabla_{\mathbf{x}} \langle T^{\left(c\right)} \rangle_{Y}\right)               \\
= R_1^{\left(c\right)} \langle T^{\left(p\right)} \rangle_{Y} - R_2^{\left(c\right)} \langle T^{\left(c\right)} \rangle_{Y} + R_3^{\left(c\right)} q^{\left(pw\right)}\left(t, \mathbf{x}\right) + R_4^{\left(c\right)} {\overline{\Pi}},}

\neweq{eq:upscale_power_flux_source}{{\overline{\Pi}} = \Pi_{\text{base}} + \frac{1}{2}\left\{\text{Erf}\left[\frac{A_1}{\phi^{\left(c\right)}} \langle T^{\left(c\right)} \rangle_{Y} + B_1\right] + 1\right\}\left(1 - \Pi_{\text{base}}\right)                   \\
- \frac{1}{2}\left\{\text{Erf}\left[\frac{A_2}{\phi^{\left(c\right)}} \langle T^{\left(c\right)} \rangle_{Y} + B_2\right] + 1\right\}.}
\end{subequations}

\noindent  where $i = p$, $c$, or $w$ refer to packing material, battery cell and cooling water pipe, respectively, $j = pc$ or $pw$ refers to the interface between packing material and battery cell, and the interface between packing material and cooling water pipes, respectively, $\phi^{\left(i\right)}$ is the volume fraction of domain $i$ in the unit cell, $|Y|$ is the area of the unit-cell domain, and $|\Gamma^{\left(j\right)}|$ is the length of interface $\Gamma^{\left(j\right)}$ in the unit-cell.  In \Cref{eq:upscaled_goven_eqs}, $\overline{\Pi}(\langle T^{\left(c\right)} \rangle_{Y}, \mathbf{x})$ is the homogenized power flux source term and ${\Pi}_{\text{base}}(\hat{\mathbf{x}})$ is the dimensionless \textit{base} power flux values, respectively; $\U^{\left(i\right)}$ are effective velocities, $\V^{\left(i\right)}$ are effective parameters corresponding to emergent terms, $\K^{\left(i\right)}$ are effective thermal conductivities, and $R_{\bm{\cdot}}^{\left(i\right)}$, with $\cdot=\{1,2,3,4\}$, and $\R_{5}^{\left(p\right)}$ are effective reaction rates, and are defined as
\begin{subequations}
\label{eq:effectivecoefficients}
\neweq{}{\U^{\left(p\right)} = \phi^{\left(p\right)}\frac{\text{Bi}^{\left(p\right)}}{|\mathcal{B}^{\left(p\right)}|} |\Gamma^{\left(pc\right)}| \langle \chiv^{\left(p\right)\left[3\right]} \rangle_{\Gamma^{\left(pc\right)}} - k^{\left(p\right)}\langle\nabla_{\xiv}\chi^{\left(p\right)\left[2\right]}\rangle_{Y},}
\neweq{}{\V^{\left(p\right)} = \frac{\phi^{\left(p\right)}}{\phi^{\left(c\right)}}\left[\phi^{\left(p\right)}\frac{\text{Bi}^{\left(p\right)}}{|\mathcal{B}^{\left(p\right)}|} |\Gamma^{\left(pc\right)}| \langle \chiv^{\left(c\right)\left[2\right]} \rangle_{\Gamma^{\left(pc\right)}} - k^{\left(p\right)}\langle \nabla_{\xiv}\chi^{\left(p\right)\left[2\right]} \rangle_{Y}\right],}
\neweq{}{\K^{\left(p\right)} = k^{\left(p\right)}\left[\phi^{\left(p\right)}\I + \langle \nabla_{\xiv}\chiv^{\left(p\right)\left[3\right]} \rangle_{Y}\right],}
\neweq{}{R_1^{\left(p\right)} = \phi^{\left(p\right)}\frac{\text{Bi}^{\left(p\right)}}{|\mathcal{B}^{\left(p\right)}|}|\Gamma^{\left(pc\right)}|\left(\frac{1}{\epsilon} - \langle\chi^{\left(c\right)\left[1\right]}\rangle_{\Gamma^{\left(pc\right)}} + \langle\chi^{\left(p\right)\left[2\right]}\rangle_{\Gamma^{\left(pc\right)}}\right),}
\neweq{}{R_2^{\left(p\right)} = \frac{\phi^{\left(p\right)}}{\phi^{\left(c\right)}}R_1^{\left(p\right)},}
\neweq{}{R_3^{\left(p\right)} = \phi^{\left(p\right)^2}\left[\frac{\mathcal{Q}|\Gamma^{\left(pw\right)}|}{|\mathcal{B}^{\left(p\right)}|\epsilon} + \frac{\text{Bi}^{\left(p\right)}}{|\mathcal{B}^{\left(p\right)}|} |\Gamma^{\left(pc\right)}| \langle \chi^{\left(p\right)\left[1\right]} \rangle_{\Gamma^{\left(pc\right)}}\right],}
\neweq{}{\R_4^{\left(p\right)} = \phi^{\left(p\right)}k^{\left(p\right)}\langle\nabla_{\xiv}\chi^{\left(p\right)\left[1\right]}\rangle_{Y}.}
\neweq{}{\U^{\left(c\right)} = \varrho \varsigma \left[\phi^{\left(c\right)}\frac{\text{Bi}^{\left(c\right)}}{|\mathcal{B}^{\left(c\right)}|} |\Gamma^{\left(pc\right)}| \langle \chiv^{\left(c\right)\left[2\right]} \rangle_{\Gamma^{\left(pc\right)}} + k^{\left(c\right)}\langle\nabla_{\xiv}\chi^{\left(c\right)\left[1\right]}\rangle_{Y}\right],}
\neweq{}{\V^{\left(c\right)} = \frac{\phi^{\left(c\right)}}{\phi^{\left(p\right)}}\varrho \varsigma \left[\phi^{\left(c\right)}\frac{\text{Bi}^{\left(c\right)}}{|\mathcal{B}^{\left(c\right)}|} |\Gamma^{\left(pc\right)}| \langle \chiv^{\left(p\right)\left[3\right]} \rangle_{\Gamma^{\left(pc\right)}} + k^{\left(c\right)}\langle\nabla_{\xiv}\chi^{\left(c\right)\left[1\right]}\rangle_{Y}\right],}
\neweq{}{\K^{\left(c\right)} = \varrho \varsigma k^{\left(c\right)}\left[\phi^{\left(c\right)}\I + \langle\nabla_{\xiv}\chiv^{\left(c\right)\left[2\right]}\rangle_{Y}\right],}
\neweq{}{R_1^{\left(c\right)} = \frac{\phi^{\left(c\right)}}{\phi^{\left(p\right)}} R_2^{\left(c\right)},}
\neweq{}{R_2^{\left(c\right)} = \phi^{\left(c\right)}\frac{\left(\text{Bi}^{\left(c\right)} \varrho \varsigma\right)}{|\mathcal{B}^{\left(c\right)}|} |\Gamma^{\left(pc\right)}| \left(\frac{1}{\epsilon} - \langle \chi^{\left(c\right)\left[1\right]} \rangle_{\Gamma^{\left(pc\right)}} + \langle \chi^{\left(p\right)\left[2\right]} \rangle_{\Gamma^{\left(pc\right)}}\right),}
\neweq{}{R_3^{\left(c\right)} = \phi^{\left(c\right)^2}\frac{\left(\text{Bi}^{\left(c\right)} \varrho  \varsigma\right)}{|\mathcal{B}^{\left(c\right)}|} |\Gamma^{\left(pc\right)}| \langle \chi^{\left(p\right)\left[1\right]} \rangle_{\Gamma^{\left(pc\right)}},}
\neweq{}{R_4^{\left(c\right)} = \phi^{\left(c\right)^2} \varrho \mathcal{R},}
\end{subequations}
\noindent with the averaging operators  in \Cref{upscaled,eq:effectivecoefficients}  defined as 
\neweqgat{eq:ave-op}{\langle \;\; \bm{\cdot}^{\left(i\right)} \rangle_{Y} \equiv \frac{1}{|Y|} \int_{\mathcal{B}^{\left(i\right)}}\left( \;\; \bm{\cdot}^{\left(i\right)}\right)\;d\xiv, \quad \langle \bm{\cdot} \rangle_{\Gamma^{\left(j\right)}} \equiv \frac{1}{|\Gamma^{\left(j\right)}|} \int_{\Gamma^{\left(j\right)}}\left(\bm{\cdot}\right)\;d\xiv.}

\noindent Furthermore, $\chi^{\left(p\right)\left[1\right]}$, $\chi^{\left(p\right)\left[2\right]}$, $\chiv^{\left(p\right)\left[3\right]}$,  $\chi^{\left(c\right)\left[1\right]}$ and $\chiv^{\left(c\right)\left[2\right]}$ are closure variables which satisfy boundary value problems to be solved in the unit cell with periodic boundary conditions~\cite{Pietrzyk2023-ou, Yao2023-zi}. A detailed formulation of all closure problems can be found in \ref{subsection:Appendix_E_Closure_Problems}. 

It is important to emphasize that  \Cref{upscaled,eq:effectivecoefficients} are valid for the following values of dimensionless numbers~\cite{Pietrzyk2023-ou}, expressed in terms of integer powers of the separation of scales parameter $\epsilon$,
\neweqgat{eq:regimes}{\text{Bi}^{\left(p\right)} \sim\mathcal{O}(1), \quad \mathcal{Q}  \sim \mathcal{O}(1), \quad \varrho \sim \mathcal{O}(1), \varsigma\sim \mathcal{O}(1),  \quad \text{Bi}^{\left(c\right)} \sim \mathcal{O}(1), \quad \mathcal{R}  \sim \mathcal{O}(\epsilon^{-1}).}
Under these conditions, the upscaled  \Cref{upscaled,eq:effectivecoefficients} are able to represent  fine-scale dynamics within upscaling errors of order $\epsilon$, as prescribed by homogenization theory; in other words, conditions \eqref{eq:regimes} ensure the accuracy and predictivity of \Cref{upscaled,eq:effectivecoefficients}.

\section{Auto-detecting and Adaptive Hybrid Coupling in 2D Domains}
\label{sec:Autodetecting_intro}

In this Section,  we consider the existence of a ``breakdown'' region of the upscaled equations presented in \Cref{sec:upscaled-govern-eqs}, denoted as ${\Omega}_{\text{break}}$, within a heterogeneous battery pack ${\Omega}_{\epsilon}$, i.e. we define  ${\Omega}_{\text{break}}$ as the region where one or more of the applicability conditions \eqref{eq:regimes} are violated (\Cref{fig:het-domain-ts}), the accuracy of \Cref{upscaled,eq:effectivecoefficients} cannot be guaranteed and the fine scale model \eqref{eq:Heat_Transfer_Eq_All} should be used instead.  

In Section \ref{sec:TS_fixed_formulation}, we first introduce the coupling strategy between the upscaled equations \Cref{upscaled} and the fine-scale  \Cref{eq:Heat_Transfer_Eq_All} which allows one to handle two-sided hybrid formulations in which the breakdown region is contained in the interior of the computational domain as shown in Figure \ref{fig:het-domain-ts}. In Section \ref{sec:ad-formulation}, we introduce the model refinement/coarsening criteria for the adaptive formulation. In Section \ref{sec:kernels}, we introduce the second-order mapping kernels for downsclaing and upscaling, and summarize the algorithm in Section \ref{sec:summary}.

\subsection{Two-sided hybrid coupling governing equations}
\label{sec:TS_fixed_formulation}

\begin{figure}
    \centering
    \includegraphics[width=0.7\textwidth]{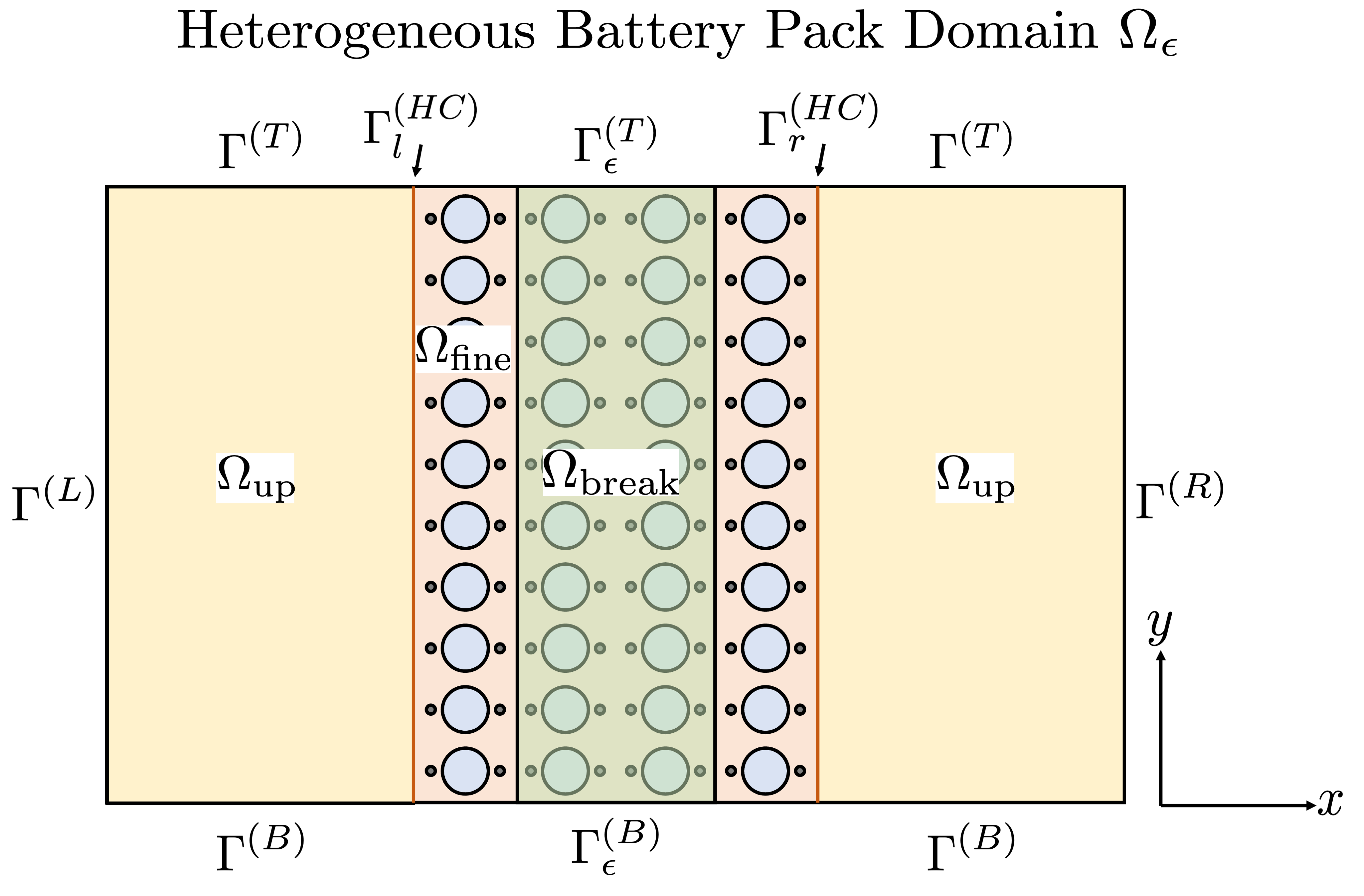}
    \caption{{Illustration of an two-sided hybrid domain consisting of both fine-scale and upscaled subdomains.}}
    \label{fig:het-domain-ts}
\end{figure}

The two-sided hybrid coupling formulation developed in this section is based on generalizing the the one-sided formulation detailed in \cite{Yao2023-zi}. 
We start by defining the fine-scale subdomain, $\Omega_{\text{fine}}$, as the slightly expanded breakdown region of the battery pack domain such that ${\Omega}_{\text{break}} \subset {\Omega}_{\text{fine}} \subset {\Omega}_{\epsilon}$  (\Cref{fig:het-domain-ts}). The upscaled subdomain, ${\Omega}_{\text{up}}$, is defined as the difference between fine-scale subdomain and battery pack domain, ${\Omega}_{\epsilon} \backslash {\Omega}_{\text{fine}}$.The boundary of ${\Omega}_{\text{up}}$ is defined by the packing edges $\Gamma^{\left(L\right)}$, $\Gamma^{\left(R\right)}$, $\Gamma^{\left(T\right)}$, and $\Gamma^{\left(B\right)}$, as well as the coupling boundaries $\Gamma_l^{\left(HC\right)}$ and $\Gamma_r^{\left(HC\right)}$. Similarly, the fine-scale subdomain $\Omega_{\text{fine}}$ is bounded by $\Gamma_\epsilon^{\left(T\right)}$ and $\Gamma_\epsilon^{\left(B\right)}$, alongside the same coupling boundaries $\Gamma_l^{\left(HC\right)}$ and $\Gamma_r^{\left(HC\right)}$ (\Cref{fig:het-domain-ts}). In hybrid simulations, quantities in ${\Omega}_{\text{fine}}$ and ${\Omega}_{\text{up}}$ are solved using fine-scale (\Cref{eq:Heat_Transfer_Eq_All}) and upscaled (\Cref{upscaled}) equations, respectively. To derive the coupling conditions between these domains, we introduce a moving average operator for fine-scale quantities as proposed in~\cite{Pietrzyk2023-ou},

\begin{subequations}
\neweq{moving_averages}{\langle \bm{\cdot} \rangle_{\mathcal{W}_{\epsilon}^{\left(p\right)}\left(\x\right)} \equiv \frac{1}{\left\vert Y \right\vert} \int_{\mathcal{W}_{\epsilon}^{\left(p\right)}\left(\x\right)}\left(\bm{\cdot}\right)\;d\y \quad \text{and} \quad \langle \bm{\cdot} \rangle_{\mathcal{W}_{\epsilon}^{\left(c\right)}\left(\x\right)} \equiv \frac{1}{\left\vert Y \right\vert} \int_{\mathcal{W}_{\epsilon}^{\left(c\right)}\left(\x\right)}\left(\bm{\cdot}\right)\;d\y,}
\noindent where
\neweq{moving_averages2_packing}{\mathcal{W}_{\epsilon}^{\left(p\right)}\left(\x\right) = \left\{ \left(x',y'\right) : x - 0.5\epsilon < x' < x + 0.5\epsilon, \; y + 0.5a\epsilon < y' < y - 0.5a\epsilon, \; \x' \in \mathcal{B}_{\epsilon}^{\left(p\right)} \right\}         \\
\text{for } \x = \left(x,y\right) \in \Omega,}
\neweq{moving_averages2_cell}{\mathcal{W}_{\epsilon}^{\left(c\right)}\left(\x\right) = \left\{ \left(x',y'\right) : x - 0.5\epsilon < x' < x + 0.5\epsilon, \; y + 0.5a\epsilon < y' < y - 0.5a\epsilon, \; \x' \in \mathcal{B}_{\epsilon}^{\left(c\right)} \right\}         \\
\text{for } \x = \left(x,y\right) \in \Omega,}
\end{subequations}

\noindent $\epsilon$ and $a$ are the length scale ratio and the aspect ratio of the unit cell defined in \Cref{eq:eps} and \Cref{eq:aspect_ratio}, respectively, and $\vert Y \vert$ is the volume of the unit cell. Following similar procedures as in \cite{Yao2023-zi}, we derive the continuity conditions for temperature and flux as 

\begin{subequations}
\label{eq:coupling-conds}
\begin{align}
&\ensmean{T^{\left(p\right)}}_{Y}(\mathbf{x}^{-}_l) = \langle T_\epsilon^{\left(p\right)} \rangle_{\mathcal{W}_{\epsilon}^{\left(p\right)}\left(\x^+_l\right)},  \quad \text{for} \quad \abs{\mathbf{x}^{+}_l - \mathbf{x}^{-}_l} \rightarrow 0, \\
&\ensmean{T^{\left(p\right)}}_{Y}(\mathbf{x}^{-}_r) = \langle T_\epsilon^{\left(p\right)} \rangle_{\mathcal{W}_{\epsilon}^{\left(p\right)}\left(\x^+_r\right)},  \quad \text{for} \quad \abs{\mathbf{x}^{+}_r - \mathbf{x}^{-}_r} \rightarrow 0, \\
& \ensmean{\mathbf{J}^{\left(p\right)}}_{Y}(\mathbf{x}^{-}_l)  \bm{\cdot} \mathbf{n}^{\left(p\right)}_{\epsilon,l} = \phi^{\left(p\right)}\ensmean{\mathbf{J}_\epsilon^{\left(p\right)}}_{\mathcal{W}_{\epsilon}^{\left(p\right)}\left(\x^+_l\right)} \bm {\cdot} \mathbf{n}^{\left(p\right)}_{\epsilon,l}, \quad \text{for} \quad \abs{\mathbf{x}_l^{+} - \mathbf{x}_l^{-}} \rightarrow 0,  \\
& \ensmean{\mathbf{J}^{\left(p\right)}}_{Y}(\mathbf{x}^{-}_r)  \bm{\cdot} \mathbf{n}^{\left(p\right)}_{\epsilon,r} = \phi^{\left(p\right)}\ensmean{\mathbf{J}_\epsilon^{\left(p\right)}}_{\mathcal{W}_{\epsilon}^{\left(p\right)}\left(\x^+_r\right)} \bm {\cdot} \mathbf{n}^{\left(p\right)}_{\epsilon,r}, \quad \text{for} \quad \abs{\mathbf{x}_r^{+} - \mathbf{x}_r^{-}} \rightarrow 0, 
 \label{eq:packing_flux_hc}
\end{align}
\end{subequations}
\noindent where {${\mathbf{J}_\epsilon^{\left(p\right)}} (\mathbf{x}^{+}_j)$ is the fine-scale fluxes of the packing materials, $\ensmean{\mathbf{J}^{\left(p\right)}}_{Y}(\mathbf{x}^{-}_j)$ is the upscaled fluxes of the packing materials}, $\x^+_j \in \Omega_{\text{fine}}$ and $\x^-_j \in \Omega_{\text{up}}$ are centroids of two unit cells defined in fine-scale and upscaled subdomains, and $j$ = $l$ or $r$ refers to the left or right coupling boundary. Since the quantities on the right side of \Cref{eq:coupling-conds} are not known, we use the Taylor approach developed in~\cite{Yao2023-zi} to approximate the unknown quantities of interest such that

\begin{subequations}
\label{eq:taylor-hc}
\begin{align}
&\langle T_\epsilon^{\left(p\right)} \rangle_{\mathcal{W}_{\epsilon}^{\left(p\right)}\left(\x^{\left(HC\right)}_j\right)} \approx \ensmean{T_\epsilon^{\left(p\right)}}_{Y_{in,j}} + \alpha^{-1} \phi^{\left(p\right)}_{out} \left[ T_\epsilon^{\left(p\right)}(\mathbf{x}^{\left(HC\right)}_j) + \pdv{T_\epsilon^{\left(p\right)}(\mathbf{x}^{\left(HC\right)}_j)}{\mathbf{x}} \left(\mathbf{x}^+_{j,out} - \mathbf{x}^{\left(HC\right)}_j\right) \right], \label{eq:taylor-hc-temp}\\
&\phi^{\left(p\right)}\ensmean{\mathbf{J}_\epsilon^{\left(p\right)}}_{\mathcal{W}_{\epsilon}^{\left(p\right)}\left(\x^{\left(HC\right)}_j\right)} \bm {\cdot} \mathbf{n}^{\left(p\right)}_{\epsilon,j} \approx \phi^{\left(p\right)}\ensmean{\mathbf{J}_\epsilon^{\left(p\right)}}_{Y_{in}} \bm{\cdot} \mathbf{n}_{\epsilon,j}^{\left(p\right)} + q^{\left(p,n\right)}_j, \label{eq:taylor-hc-flux}\\
& \mathbf{J}_\epsilon^{\left(p\right)}(\mathbf{x}^{\left(HC\right)}_j) \bm{\cdot} \mathbf{n}_{\epsilon,j}^{\left(p\right)} \approx {\alpha}{\left(\phi^{\left(p\right)} \phi^{\left(p\right)}_{out}\right)}^{-1} q^{\left(p,n\right)}_j,
\end{align}
\noindent where 
\begin{align}
& \ensmean{T_\epsilon^{\left(p\right)}}_{Y_{in,j}} = \ddfrac{1}{\abs{Y}} \int_{\mathcal{B}^{\left(p\right)}_{in,j}} T_\epsilon^{\left(p\right)}(\mathbf{y}) \ \dd\mathbf{y}, \\
& \ensmean{\mathbf{J}_\epsilon^{\left(p\right)}}_{Y_{in,j}} = \ddfrac{1}{\abs{Y}} \int_{\mathcal{B}^{\left(p\right)}_{in,j}} \mathbf{J}_\epsilon^{\left(p\right)}(\mathbf{y}) \ \dd\mathbf{y}, 
\end{align}
\neweq{eq:packing_in}{\mathcal{B}_{in,j}^{\left(p\right)} = \left\{ \left(x',y'\right) : x^{\left(HC\right)}_{j} < x' < x^{\left(HC\right)}_{j} + 0.5\epsilon, \; y^{\left(HC\right)}_{j} - 0.5a\epsilon < y' < y^{\left(HC\right)}_{j} + 0.5a\epsilon, \; \x' \in \mathcal{B}_{\epsilon}^{\left(p\right)} \right\}         \\
\text{for } \x = \left(x,y\right) \in \Omega,}
\neweq{}{\phi^{\left(p\right)}_{out} = \frac{\abs{Y}\phi^{(p)} -  \int_{\mathcal{B}^{\left(p\right)}_{in}} \mathbf{1} \ \dd\mathbf{y}}{\abs{Y_{out}}}.}
\end{subequations}

\noindent In \Cref{eq:taylor-hc-flux}, $q^{\left(p,n\right)}_j$ is the unresolved heat flux of the packing material across the coupling boundary $j$ at iteration $n$. {For numerical implementation, the equations were discretized with the backward Euler method to avoid time-step constraints. Both sets of equations were solved using an open-source finite element code, FEniCS~\cite{Logg2012-ul}.}

\subsection{Automatic detecting and adaptive formulation}
\label{sec:ad-formulation}

As demonstrated by~\cite{Yao2023-zi}, the upscaled simulations become less accurate when the magnitude of the dimensionless numbers violates the applicability regimes prescribed by homogenization theory. The magnitude of the dimensionless numbers controlling the accuracy of upscaled models can be space and time dependent, due, e.g. to aging effects, manufacturing defects etc. Hence, the ability to adaptively identify the space-time regions (i.e. the location and boundaries of the breakdown region) where upscaled models loose predictivity, and hybridization needs to occur, is necessary to maintain accuracy throughout the simulation. Here, we present an automated method for the detection and (space-time) adaptation of these regions. This method computes and estimates the extent of the breakdown region based on a set of dimensionless numbers. Each dimensionless number $i$ is represented as $DN_{i}$, with its applicable range defined as $DN_{i,app}$. The left boundary of $DN_{i}$ is defined as
\begin{align}
    \x^{\min}_{DN,i} = \left[ \argmin_{\x} \left( \left\vert \ddfrac{DN_{i}}{DN_{i,app}} - 1 - \alpha_1  \right\vert \right) \right] - \alpha_2\epsilon,
\end{align}

\noindent where $\alpha_1$ is the tolerance factor and $\alpha_2$ is a factor that controls the additional volume of non-breakdown region to be included in the fine-scale subdomain. Large $\alpha_1$ and $\alpha_2$ will result in a larger fine-scale subdomain, therefore increasing the computational cost. In our simulations, we use $\alpha_1 = 0.01$ and $\alpha_2=1.5$. To note, $\alpha_2=1.5$ is the minimum factor required for hybrid simulations to be accurate, as demonstrated by~\cite{Yao2023-zi}. A similar approach is used to calculate the location of right boundary of $DN_i$ such that 
\begin{align}
    \x^{\max}_{DN,i} = \left[ \argmax_{\x} \left( \left\vert \ddfrac{DN_{i}}{DN_{i,app}} - 1 - \alpha_1  \right\vert \right) \right] + \alpha_2\epsilon.
\end{align}

In the problem of thermal runaway in a battery pack, there are numerous dimensionless numbers. For a system with multiple dimensionless numbers, the breakdown boundaries are defined as
\begin{align}
\label{eq:AD_for_all_DNs}
    \begin{cases}
    \x^{\min}_{DN} = \min \left( \x^{\min}_{DN,1}, \x^{\min}_{DN,2} \cdots \x^{\min}_{DN,m} \right), \\
    \x^{\max}_{DN} = \max \left( \x^{\max}_{DN,1}, \x^{\max}_{DN,2} \cdots \x^{\max}_{DN,m} \right),
    \end{cases}
\end{align}
\noindent where $m$ is the number of dimensionless numbers. 

\subsection{Second-order accurate mapping kernels}
\label{sec:kernels}

In the automatic-detecting and adaptive hybrid method, the spatial subdomains in which fine-scale or continuum-scale equations are solved evolve with time, as a result of the dynamic values of the dimensionless numbers (which, in turn, reflect  changes in the magnitude of the dependent variables spatial gradients); as such, mapping values between fine-scale and upscaled subdomains is essential, since, as time evolves a continuum-scale region may require  pore-scale resolution, and viceversa. In this section, we define a downscaling kernel, used to interpolate values from upscaled to fine-scale subdomains, and an upscaling kernel to map values from fine-scale to upscaled subdomains. To ensure that errors remain within acceptable upscaling error margins, it is necessary to formulate kernels that match the accuracy of the upscaling techniques. We will discuss the detailed derivation of these kernels in this section.

\subsubsection{Downscaling kernel: expanding the fine-scale subdomain and reducing the upscaled subdomains}
\label{sec:downscale_ker}

\begin{figure}
    \centering
    \subfloat[]{\includegraphics[width=\textwidth]{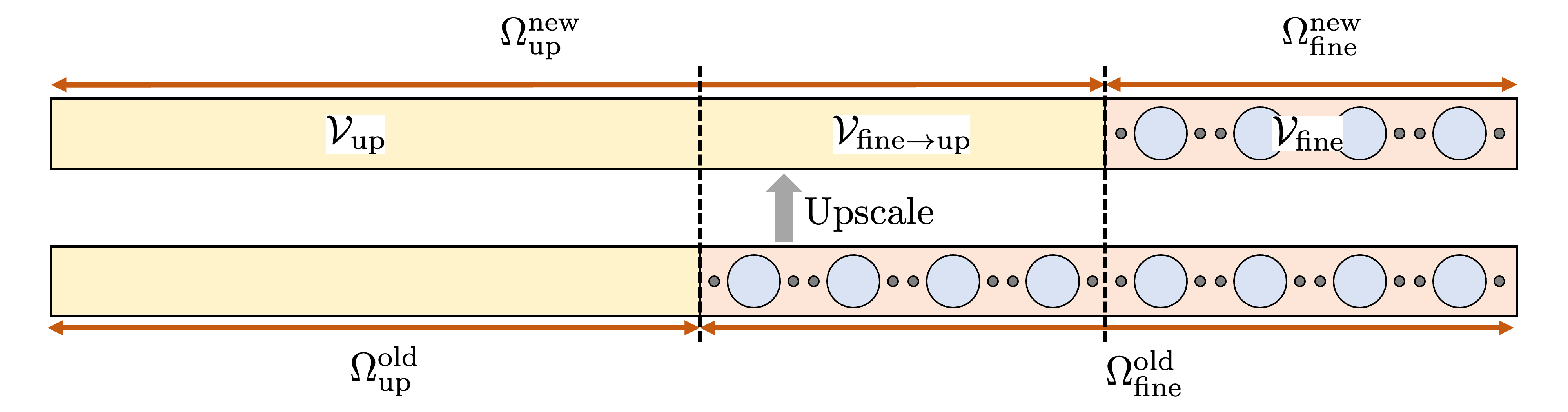}
    \label{fig:downscale_illus}}
    
    \subfloat[]{\includegraphics[width=\textwidth]{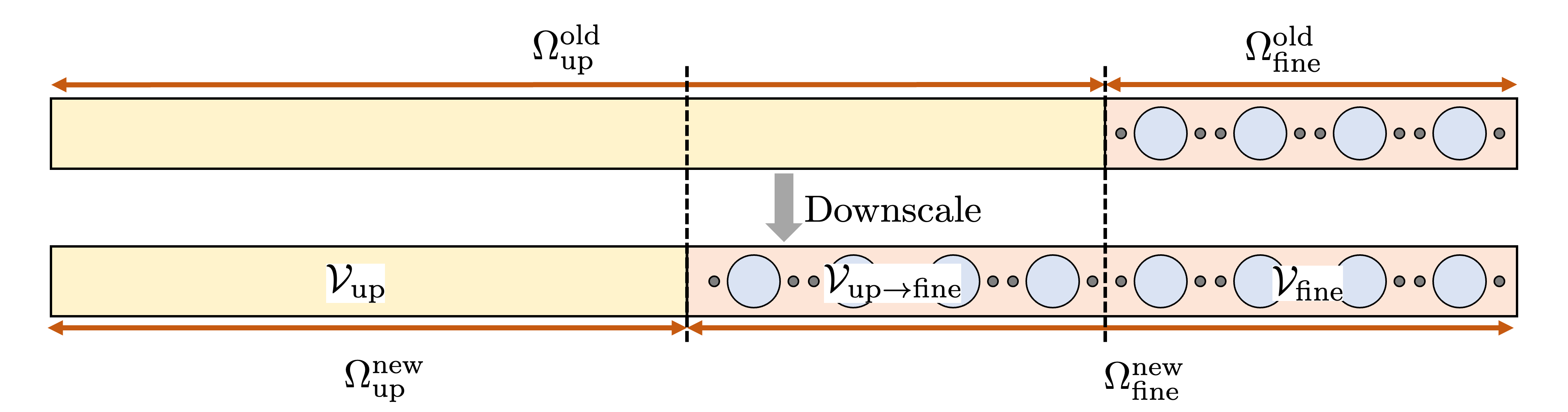}
    \label{fig:upscale_illus}}     
    
    \subfloat[]{\includegraphics[width=0.9\textwidth]{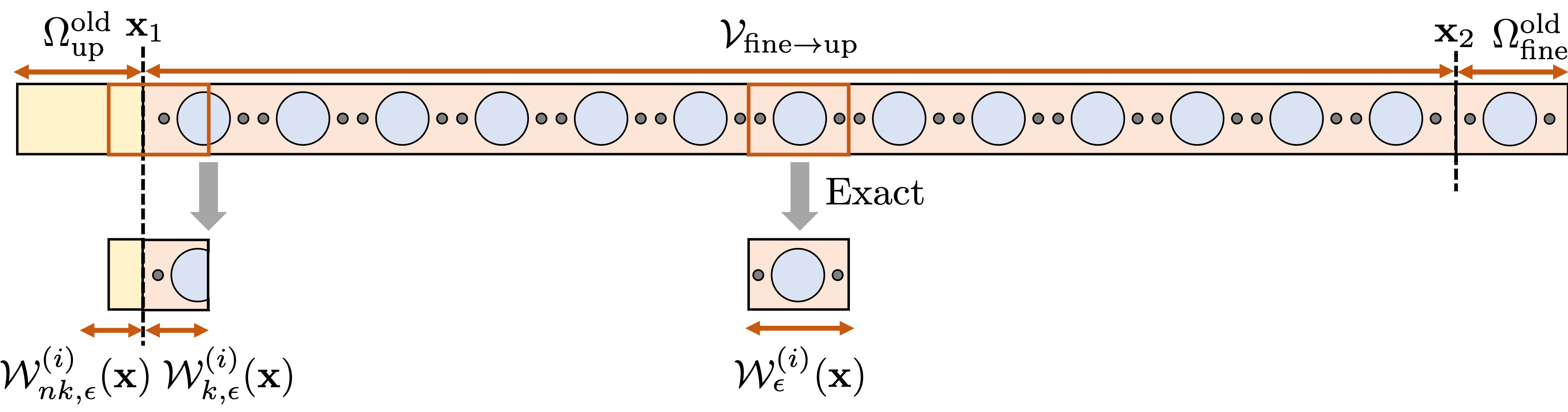}
    \label{fig:upscale_sub_illus}}  
     
     \caption{Schematic illustration of (a) expanded fine-scale subdomain (downscaled) (b) reduced fine-scale subdomain (upscaled) from time $n$ to $n+1$, and (c) known and unknown subdomains $\mathcal{W}_{k,\epsilon}^{\left(i\right)}$ and $\mathcal{W}_{nk,\epsilon}^{\left(i\right)}$ in $\mathcal{V}_{\text{fine}\rightarrow \text{up}}$.}
     \label{fig:scale_illus}
\end{figure}

As time marches from $t=n$ to $t=n+1$, the breakdown region expands, resulting in an increase in the size of the fine-scale subdomain and a decrease in the size of the upscaled subdomain. To represent the expansion of the breakdown region, we define two computational domains $\Omega^{\text{old}} = \Omega_{\text{fine}}^{\text{old}} \cup \Omega_{\text{up}}^{\text{old}}$ and $\Omega^{\text{new}} = \Omega_{\text{fine}}^{\text{new}} \cup \Omega_{\text{up}}^{\text{new}}$ (\Cref{fig:downscale_illus}) where the superscript ``old" and ``new" refer to the corresponding domain or subdomain before and after the expansion of the breakdown region.

To differentiate the quantities of interest in the two domains ($\Omega^{\text{old}}$ and $\Omega^{\text{new}}$), we use $T^{\left(i\right)}_{\epsilon}\left(\x,t\right)$ and $\langle T^{\left(i\right)} \rangle_{Y}\left(\x, t\right)$ to denote the fine-scale and upscaled temperatures in the original domain $\x \in \Omega^{\text{old}}$, and $\Theta^{\left(i\right)}_{\epsilon}\left(\x,t\right)$ and $\langle \Theta^{\left(i\right)} \rangle_{Y}\left(\x, t\right)$ to denote the corresponding temperatures in the new domain $\x \in \Omega^{\text{new}}$. To march $\Theta^{\left(i\right)}_{\epsilon}$ and $\langle \Theta^{\left(i\right)} \rangle_{Y}$ from $t=n$ to $t=n+1$, these quantities at $t=n$ are necessary. However, at time $t=n$, fine-scale and upscaled equations are only solved in the original domain $\Omega^{\text{old}}$, therefore, only $T^{\left(i\right)}_{\epsilon}$ and $\langle T^{\left(i\right)} \rangle_{Y}$ are available only in a subset of the necessary domain. As such,  accurate mapping kernels from $(T^{\left(i\right)}_{\epsilon}, \langle T^{\left(i\right)} \rangle_{Y})$ to $(\Theta^{\left(i\right)}_{\epsilon}, \langle \Theta^{\left(i\right)} \rangle_{Y})$ are therefore necessary.

We can divide the new domain $\Omega^{\text{new}}$ into three different subdomains (\Cref{fig:downscale_illus}): (1) $\mathcal{V}_{\text{up}}=\Omega_{\text{up}}^{\text{old}} \cap \Omega_{\text{up}}^{\text{new}}$, (2) $\mathcal{V}_{\text{up}\rightarrow \text{fine}} = \Omega_{\text{up}}^{\text{old}} \backslash \Omega_{\text{up}}^{\text{new}}$ or $\Omega_{\text{fine}}^{\text{new}} \backslash \Omega_{\text{fine}}^{\text{old}}$ and (3) $\mathcal{V}_{\text{fine}}=\Omega_{\text{fine}}^{\text{old}} \cap \Omega_{\text{fine}}^{\text{new}}$. For the downscaling kernel, each subdomain is treated differently to map the quantities of interest $\Theta^{\left(i\right)}_{\epsilon}$ and $\langle \Theta^{\left(i\right)} \rangle_{Y}$ from $T^{\left(i\right)}_{\epsilon}$ and $ \langle T^{\left(i\right)} \rangle_{Y}$ at time $t=n$.

\noindent\textbf{Subdomain $\mathcal{V}_{\text{up}}$}.  $\mathcal{V}_{\text{up}}$ is the common upscaled subdomain existing  both in $\Omega^{\text{old}}$ and $\Omega^{\text{new}}$, satisfying $\x \in \mathcal{V}_{\text{up}} \subset \Omega_{\text{up}}^{\text{old}}$. As such, $\ensmean{\Theta^{\left(i\right)}}_Y$ is well defined for $\x \in \mathcal{V}_{\text{up}}$, allowing one-to-one mapping from 
$\ensmean{T^{\left(i\right)}}_Y$  to $\ensmean{\Theta^{\left(i\right)}}_Y$. The mapping is therefore defined as
\begin{align}
     \ensmean{\Theta^{\left(i\right)}}_Y (\x) = \ensmean{T^{\left(i\right)}}_Y (\x), \quad \x \in \mathcal{V}_{\text{up}} \subset \Omega_{\text{up}}^{\text{old}}.
\end{align}
\textbf{Subdomain $\mathcal{V}_{\text{fine}}$}. Similarly to before, $\mathcal{V}_{\text{fine}}$ is the common fine-scale subdomain existing in both $\Omega^{\text{old}}$ and $\Omega^{\text{new}}$, satisfying $\x \in \mathcal{V}_{\text{fine}} \subseteq \Omega_{\text{fine}}^{\text{old}}$. This enables a one-to-one mapping from $T^{\left(i\right)}_\epsilon$ to $\Theta^{\left(i\right)}_\epsilon$ such that
\begin{align}
     \Theta^{\left(i\right)}_\epsilon(\x) = T^{\left(i\right)}_\epsilon(\x), \quad \x \in \mathcal{V}_{\text{fine}} \subseteq \Omega_{\text{fine}}^{\text{old}}.
\end{align}
\textbf{Subdomain $\mathcal{V}_{\text{up} \rightarrow \text{fine}}$}. $\mathcal{V}_{\text{up} \rightarrow \text{fine}}$ is the upscaled subdomain in $\Omega^{\text{old}}$ where model granularity is increased, i.e. in $\Omega^{\text{new}}$ the model must be switched from upscaled to fine-scale.  As such, only $\ensmean{T^{\left(i\right)}}_Y$ are defined in $\mathcal{V}_{\text{up} \rightarrow \text{fine}}$ at timestep $n$. Because in $\Omega^{\text{new}}$, fine-scale equations need to be solved in $\mathcal{V}_{\text{up} \rightarrow \text{fine}}$ and marched in time from timestep $n$ to timestep $n+1$,  $\ensmean{T^{\left(i\right)}}_Y$ for $t=n$ must be mapped onto $\Theta^{\left(i\right)}_\epsilon$ for the same timestep. Here, we define the interpolation functions as
\begin{subequations}
\label{eq:downscale_interp}
\begin{align}
& \ensmean{\Theta_\epsilon^{\left(i\right)}}_{\mathcal{W}_{\epsilon}^{\left(i\right)}\left(\x\right)} = \mathcal{I}\left(\ensmean{T^{\left(i\right)}}_Y (\y)\right), \quad \x \in \mathcal{V}_{\text{up}\rightarrow \text{fine}}, \y \in \Omega_{\text{up}}^{\text{old}} \label{eq:downscale_interp_a} 
 \\
& \Theta^{\left(i,\right)}_\epsilon (\x) =\mathcal{M}\left(\ensmean{\Theta_\epsilon^{\left(i\right)}}_{\mathcal{W}_{\epsilon}^{\left(i\right)}\left(\x\right)}\right),  \quad \x \in \mathcal{V}_{\text{up} \rightarrow \text{fine}}, \label{eq:downscale_interp_b}
\end{align}
\end{subequations}
where $\mathcal{I}\left(\cdot\right)$ is a second-order-accurate scheme to map quantities from a coarser grid ($\Omega_{\text{up}}^{\text{old}}$) to a finer grid ($\mathcal{V}_{\text{up} \rightarrow \text{fine}}$), and $\mathcal{M}\left(\cdot\right)$ is another second-order accurate interpolation function to map upscaled quantities to fine-scale quantities on the same grid. Typical examples of $\mathcal{I}\left(\cdot\right)$ are polynomial interpolation used in the Finite Element Method or bilinear interpolation used in a structured grid. 

To derive the interpolation function $\mathcal{M}\left(\cdot\right)$ in \Cref{eq:downscale_interp_b} to map upscaled temperatures to fine-scale temperatures, we consider the definition of $\ensmean{\Theta_\epsilon^{\left(i\right)}}_{\mathcal{W}_{\epsilon}^{\left(i\right)}\left(\x\right)}$ as 
\begin{align}
    \label{eq:fine-up-equi}
    \ensmean{\Theta_\epsilon^{\left(i\right)}}_{\mathcal{W}_{\epsilon}^{\left(i\right)}\left(\x\right)} = \frac{1}{\left\vert Y \right\vert} \int_{\mathcal{W}_{\epsilon}^{\left(i\right)}\left(\x\right)}\Theta_\epsilon^{\left(i\right)}\left(\y\right)\;d\y,
\end{align}
where $\abs{Y} = \nicefrac{\abs{\mathcal{B}_\epsilon^{\left(i\right)}}}{ \phi^{\left(i\right)}}$, and $\phi^{\left(i\right)}$ is the volume fraction of domain $i$ (either packing material or battery cells) in the unit cell. Applying Taylor expansion to $\Theta_\epsilon^{\left(i\right)}\left(\y\right)$ at location $\x_c$ where $\x_c$ is the centroid of the moving averaging window $\mathcal{W}_{\epsilon}^{\left(i\right)}\left(\x\right)$, we obtain 

\begin{align}
    \label{eq:taylor_expan}
    \Theta_\epsilon^{\left(i\right)}\left(\y\right) = \Theta_\epsilon^{\left(i\right)} \bigg\vert_{\y = \x_c} + \grad{\Theta_\epsilon^{\left(i\right)}}\bigg\vert_{\y = \x_c}^T \left(\y - \x_c\right) +  \frac{1}{2} \left(\y - \x_c\right)^T \mathbf{H}_{\Theta_\epsilon^{\left(i\right)}}\bigg\vert_{\y = \x_c} \left(\y - \x_c\right) + \mathcal{O}\left( \left(\y - \x_c\right)^3\right),
\end{align}

\noindent where $\mathbf{H}_{\Theta_\epsilon^{\left(i\right)}}$ is the Hessian matrix of $\Theta_\epsilon^{\left(i\right)}$. By substituting~\Cref{eq:taylor_expan} into~\Cref{eq:fine-up-equi}, we obtain
\begin{align}
\label{eq:taylor_expan_long}
\frac{1}{\left\vert Y \right\vert} \int_{\mathcal{W}_{\epsilon}^{\left(i\right)}\left(\x\right)} \Theta_\epsilon^{\left(i\right)}\left(\y\right)\dd\y = \frac{1}{\left\vert Y \right\vert} \Bigg[ &\int_{\mathcal{W}_{\epsilon}^{\left(i\right)}\left(\x\right)} \Theta_\epsilon^{\left(i\right)} \bigg\vert_{\y = \x_c}  + \grad{\Theta_\epsilon^{\left(i\right)}}\bigg\vert_{\y = \x_c}^T \left(\y - \x_c\right) \nonumber \\ 
&+  \frac{1}{2} \left(\y - \x_c\right)^T \mathbf{H}_{\Theta_\epsilon^{\left(i\right)}}\bigg\vert_{\y = \x_c} \left(\y - \x_c\right) + \mathcal{O}\left( \left(\y - \x_c\right)^3\right) \ \dd\y \Bigg].
\end{align}

\noindent \Cref{eq:taylor_expan_long} can be written as
\begin{align}
    \label{eq:taylor_expan_long_rewritten}
    \frac{1}{\left\vert Y \right\vert} \int_{\mathcal{W}_{\epsilon}^{\left(i\right)}\left(\x\right)}\Theta_\epsilon^{\left(i\right)}\left(\y\right) \dd\y &= \frac{1}{\left\vert Y \right\vert} \Theta_\epsilon^{\left(i\right)} \bigg\vert_{\y = \x_c} \int_{\mathcal{W}_{\epsilon}^{\left(i\right)}\left(\x\right)} \mathbf{1} \ \dd\y \\ \nonumber
    &+ \frac{1}{\left\vert Y \right\vert} \grad\Theta_\epsilon^{\left(i\right)} \bigg\vert_{\y = \x_c}^T \left[\int_{\mathcal{W}_{\epsilon}^{\left(i\right)}\left(\x\right)} \y \ \dd\y  - \x_c \int_{\mathcal{W}_{\epsilon}^{\left(i\right)}\left(\x\right)} \mathbf{1} \ \dd\y \right] \\ \nonumber
    &+ \frac{1}{\left\vert Y \right\vert} \left[\int_{\mathcal{W}_{\epsilon}^{\left(i\right)}\left(\x\right)} \frac{1}{2} \left(\y - \x_c\right)^T \mathbf{H}_{\Theta_\epsilon^{\left(i\right)}}\bigg\vert_{\y = \x_c} \left(\y - \x_c\right) + \mathcal{O}\left( \left(\y - \x_c\right)^3\right) \ \dd\y \right].
\end{align} 

\noindent Using the definition of the centroid $\x_c$ and the volume of unit cell $\left\vert Y \right\vert$ as 
\begin{subequations}
\begin{align}
    \x_c &= \frac{1}{\phi^{\left(i\right)} \left\vert Y \right\vert} \int_{\mathcal{W}_{\epsilon} ^{\left(i\right)}\left(\x\right)} \y \ \dd\y, \\ 
    \left\vert Y \right\vert &= \frac{1}{\phi^{\left(i\right)}} \int_{\mathcal{W}_{\epsilon}^{\left(i\right)}\left(\x\right)} \mathbf{1} \ \dd\y,
\end{align}
\end{subequations}

\noindent the second term in bracket of \Cref{eq:taylor_expan_long_rewritten} vanishes and simplifies to
\begin{align}
    \label{eq:2nd_order_upscaled}
    \frac{1}{\left\vert Y \right\vert} \int_{\mathcal{W}_{\epsilon}^{\left(i\right)}\left(\x\right)}\Theta_\epsilon^{\left(i\right)}\left(\y\right) \dd\y &= \phi^{\left(i\right)} \Theta_\epsilon^{\left(i\right)} \bigg\vert_{\y = \x_c} + \int_{\mathcal{W}_{\epsilon}^{\left(i\right)}\left(\x\right)}  \mathcal{O}\left(\left(\y-\x_c\right)^2\right) \ \dd\y. 
\end{align}    

\noindent Finally,~\Cref{eq:2nd_order_upscaled} is substituted to~\Cref{eq:downscale_interp_b} such that

\begin{subequations}
\begin{align}
&\ensmean{\Theta_\epsilon^{\left(i\right)}}_{\mathcal{W}_{\epsilon}^{\left(i\right)}\left(\x\right)} = \phi^{\left(i\right)} \Theta_\epsilon^{\left(i\right)} \bigg\vert_{\y = \x_c} + \int_{\mathcal{W}_{\epsilon}^{\left(i\right)}\left(\x\right)}  \mathcal{O}\left(\left(\y-\x_c\right)^2\right) \ \dd\y, \text{  or} \\
&\Theta_\epsilon^{\left(i\right)}\left(\x\right) \approx \left(\phi^{\left(i\right)}\right)^{-1} \ensmean{\Theta_\epsilon^{\left(i\right)}}_{\mathcal{W}_{\epsilon}^{\left(i\right)}\left(\x\right)}.
\end{align}
\end{subequations}

In summary, the second-order downscaling kernel is defined as 
\begin{align}
\label{eq:upscaled_ker_complete}
\begin{cases}
     \ensmean{\Theta^{\left(i\right)}}_Y (\x) = \ensmean{T^{\left(i\right)}}_Y (\x), \quad \x \in \mathcal{V}_{\text{up}} \subset \Omega_{\text{up}}^{\text{old}}, \\
     \ensmean{\Theta_\epsilon^{\left(i\right)}}_{\mathcal{W}_{\epsilon}^{\left(i\right)}\left(\x\right)} = \mathcal{I}\left(\ensmean{T^{\left(i\right)}}_Y (\y)\right), \quad \x \in \mathcal{V}_{\text{up}\rightarrow \text{fine}}, \y \in \Omega_{\text{up}}^{\text{old}} \\
    \Theta_\epsilon^{\left(i\right)}\left(\x\right) = \left(\phi^{\left(i\right)}\right)^{-1} \ensmean{\Theta_\epsilon^{\left(i\right)}}_{\mathcal{W}_{\epsilon}^{\left(i\right)}\left(\x\right)},  \quad \x \in \mathcal{V}_{\text{up} \rightarrow \text{fine}},\\
    \Theta^{\left(i\right)}_\epsilon(\x) = T^{\left(i\right)}_\epsilon(\x), \quad \x \in \mathcal{V}_{\text{fine}} \subseteq \Omega_{\text{fine}}^{\text{old}}.
\end{cases}
\end{align}

\subsubsection{Upscaling kernel: reducing the fine-scale subdomain and expanding the upscaled subdomains}
\label{sec:upscale_ker}

While the downscaling kernel addresses scenarios in hybrid simulations where the fine-scale subdomain expands and the upscaled subdomain contracts,  it is also possible to encounter situations where the breakdown region shrinks over time. In such cases, the fine-scale subdomain decreases in size, while the upscaled subdomain correspondingly expands. These circumstances require a different kernel, known as the upscaling kernel, to effectively manage the changes in the simulations.

Following the approach in~\Cref{sec:downscale_ker}, we define the computational domains at time $t=n$ and $t=n+1$ as $\Omega^{\text{old}} = \Omega_{\text{fine}}^{\text{old}} \cup \Omega_{\text{up}}^{\text{old}}$ and $\Omega^{\text{new}} = \Omega_{\text{fine}}^{\text{new}} \cup \Omega_{\text{up}}^{\text{new}}$, respectively (\Cref{fig:upscale_illus}). Again, we use $T^{\left(i\right)}_{\epsilon}\left(\x,t\right)$ and $\langle T^{\left(i\right)} \rangle_{Y}\left(\x, t\right)$ to denote the fine-scale and upscaled temperatures in the original domain $\x \in \Omega^{\text{old}}$ and $\Theta^{\left(i\right)}_{\epsilon}\left(\x,t\right)$ and $\langle \Theta^{\left(i\right)} \rangle_{Y}\left(\x, t\right)$ to denote the corresponding temperatures in the new domain $\x \in \Omega^{\text{new}}$. To derive the upscaled kernel, we further define three subdomains based on the relationship between $\Omega^{\text{old}}$ and $\Omega^{\text{new}}$ (\Cref{fig:upscale_illus}): (1) $\mathcal{V}_{\text{up}}=\Omega_{\text{up}}^{\text{old}} \cap \Omega_{\text{up}}^{\text{new}}$, (2) $\mathcal{V}_{\text{fine}\rightarrow \text{up}}=\Omega_{\text{up}}^{\text{new}} \backslash \Omega_{\text{up}}^{\text{old}}$ or $\Omega_{\text{fine}}^{\text{old}} \backslash \Omega_{\text{fine}}^{\text{new}}$ and (3) $\mathcal{V}_{\text{fine}}=\Omega_{\text{fine}}^{\text{old}} \cap \Omega_{\text{fine}}^{\text{new}}$. 

\noindent \textbf{Subdomains $\mathcal{V}_{\text{up}}$ and  $\mathcal{V}_{\text{fine}}$.} For both domains, one-to-one mapping approaches can be adopted:
\begin{subequations}
\begin{align}
   & \ensmean{\Theta^{\left(i\right)}}_Y (\x) = \ensmean{T^{\left(i\right)}}_Y (\x), \quad \x \in \mathcal{V}_{\text{up}} \subset \Omega_{\text{up}}^{\text{old}}, \\
  &  \Theta^{\left(i\right)}_\epsilon(\x) = T^{\left(i\right)}_\epsilon(\x), \quad \x \in \mathcal{V}_{\text{fine}} \subseteq \Omega_{\text{fine}}^{\text{old}}.
\end{align}
\end{subequations}

\textbf{Subdomain $\mathcal{V}_{\text{fine} \rightarrow \text{up}}$.} $T_\epsilon^{\left(i\right)}$ is defined at $t=n$ since the fine-scale equations are solved in  $\Omega^{\text{old}}$. However, $\ensmean{\Theta^{\left(i\right)}}_Y$  is required  since the upscaled governing equations will be solved in $\Omega^{\text{new}}$ at time $t=n+1$. Therefore, for the subdomain $\mathcal{V}_{\text{fine} \rightarrow \text{up}}$,  $T_\epsilon^{\left(i\right)}$ must be mapped onto $\ensmean{\Theta^{\left(i\right)}}_Y$ using the definition of moving average where
\begin{align}
    & \ensmean{\Theta^{\left(i\right)}}_Y \left(\x\right) = \left\langle T_\epsilon^{\left(i\right)}\right\rangle_{\mathcal{W}_{\epsilon}^{\left(i\right)}\left(\x\right)} , \quad \x \in \mathcal{V}_{\text{fine}\rightarrow \text{up}}. \label{eq:upscale_kernel_averaging}
\end{align}

Yet, as $\x$ moves closer to the boundary between the subdomain $\mathcal{V}_{\text{fine}\rightarrow \text{up}}$ and $\Omega_{\text{up}}^{\text{old}}$ (\Cref{fig:upscale_sub_illus}), $T_\epsilon^{\left(i\right)}$ values are not available.
To illustrate this, we consider a subdomain $\mathcal{V}_{\text{fine} \rightarrow \text{up}}$ between two boundaries $\x_1$ (left) and $\x_2$ (right), with $\x_1$  the boundary in contact with $\Omega_{\text{up}}^{\text{old}}$, and  $\x_2$ the boundary in contact with $\Omega_{\text{fine}}^{\text{old}}$.  Based on the definition of $\mathcal{W}_{\epsilon}^{\left(i\right)}\left(\x\right)$ in \Cref{moving_averages2_packing,moving_averages2_cell}, $\ensmean{\Theta^{\left(i\right)}}_Y$ can be computed exactly as long as $\mathcal{W}_{\epsilon}^{\left(i\right)}\left(\x\right) \subset \mathcal{V}_{\text{fine}\rightarrow \text{up}}$ is satisfied (i.e., $\x_1 + \epsilon/2 \le \x \le \x_2$, which ensures that the averaging domain is fully contained in $\mathcal{V}_{\text{fine}\rightarrow \text{up}}$). For $\x_1 \leq \x < \x_1 + \epsilon/2$ where $\mathcal{W}_{\epsilon}^{\left(i\right)}\left(\x\right) \not \subset \mathcal{V}_{\text{fine}\rightarrow \text{up}}$, a second-order approximation method based on Taylor expansion is derived to compute $\ensmean{\Theta^{\left(i\right)}}_Y$ (\Cref{fig:upscale_sub_illus}). 

Let us consider a moving averaging window such that $\mathcal{W}_{\epsilon}^{\left(i\right)}\left(\x\right) \not \subset \mathcal{V}_{\text{fine}\rightarrow \text{up}}$. We divide the moving averaging window into two regions, a  region $\mathcal{W}_{\epsilon, \text{k}}^{\left(i\right)}\left(\x\right) = \mathcal{W}_{\epsilon}^{\left(i\right)} \cap \mathcal{V}_{\text{fine}\rightarrow \text{up}}$, in which $T_\epsilon^{\left(i\right)}$ is known, 
and a region $\mathcal{W}_{\epsilon,\text{nk}}^{\left(i\right)}\left(\x\right) = \mathcal{W}_{\epsilon}^{\left(i\right)} \cap \Omega_{\text{up}}^{\text{old}}$ in which $T_\epsilon^{\left(i\right)}$ is unknown. As such, $\left\langle T_\epsilon^{\left(i\right)}\right\rangle_{\mathcal{W}_{\epsilon}^{\left(i\right)}\left(\x\right)}$ can be expressed as
\begin{align}
    \left\langle T_\epsilon^{\left(i\right)}\right\rangle_{\mathcal{W}_{\epsilon}^{\left(i\right)}\left(\x\right)} = \frac{1}{\left\vert Y \right\vert} \int_{\mathcal{W}_{\epsilon,\text{k}}^{\left(i\right)}\left(\x\right)} T_\epsilon^{\left(i\right)} \left(
    \y\right)\;d\y + \frac{1}{\left\vert Y \right\vert} \int_{\mathcal{W}_{\epsilon,\text{nk}}^{\left(i\right)}\left(\x\right)} T_\epsilon^{\left(i\right)} \left(
    \y\right)\;d\y. \label{eq:upscale_uk}
\end{align}
To approximate the second term on the right-hand-side of \Cref{eq:upscale_uk}, we expand $T_\epsilon^{\left(i\right)} \left(\y\right)$ at $\x_{\text{k}}$ with $\x_{\text{k}}$ the nearest location where $T_\epsilon^{\left(i\right)} \left(\y\right)$ is defined in $\mathcal{W}_{\epsilon,\text{k}}^{\left(i\right)}\left(\x\right)$. As such, we can express the integral as 
\begin{align}
\frac{1}{\left\vert Y \right\vert} \int_{\mathcal{W}_{\epsilon,\text{nk}}^{\left(i\right)}\left(\x\right)} T_\epsilon^{\left(i\right)} \left(
    \y\right)\dd\y = \frac{1}{\left\vert Y \right\vert} \int_{\mathcal{W}_{\epsilon,\text{nk}}^{\left(i\right)}\left(\x\right)} \left[T_\epsilon^{\left(i\right)} \bigg\vert_{\y = \x_k} + \grad{T_\epsilon^{\left(i\right)}} \bigg\vert_{\x_{\text{k}}}^T \left(\y - \x_{\text{k}}\right) + \mathcal{O}\left(\left(\y - \x_{\text{k}} \right)^2\right) \right]\ \dd\y.
\end{align}
By defining the centroid of $\mathcal{W}_{\epsilon,\text{nk}}^{\left(i\right)}\left(\x\right)$ as 
\begin{align}
    \x_{c,\text{nk}} &= \frac{1}{\phi^{\left(i\right)} \left\vert Y_{\text{nk}} \right\vert}  \int_{\mathcal{W}_{\epsilon, \text{nk}}^{\left(i\right)}\left(\x\right)} \y \ \dd\y,
\end{align}

\noindent the expression can be simplified to
\begin{align}
\frac{1}{\left\vert Y \right\vert} \int_{\mathcal{W}_{\epsilon,\text{nk}}^{\left(i\right)}\left(\x\right)} T_\epsilon^{\left(i\right)} \left(
    \y\right)\dd\y &= \beta \phi^{\left(i\right)} \left[ T_\epsilon^{\left(i\right)} \bigg\vert_{\y = \x_k} + \grad{T_\epsilon^{\left(i\right)}}\bigg\vert_{\y = \x_k}^T  \left(\x_{c,\text{nk}} - \x_{\text{k}}\right) \right] \nonumber \\
    &+ \frac{1}{\left\vert Y \right\vert} \int_{\mathcal{W}_{\epsilon,\text{nk}}^{\left(i\right)}\left(\x\right)}  \mathcal{O}\left(\left(\y - \x_{\text{k}} \right)^2\right) \ \dd\y, \label{eq:simplified}
\end{align}
where $\beta = \ddfrac{\abs{Y_{\text{nk}}}}{\abs{Y}}$, $\abs{Y_{\text{nk}}}$ is the volume of the unknown region $\mathcal{W}_{\epsilon,\text{nk}}^{\left(i\right)}\left(\x\right)$ in the averaging window, and $\norm{\x_{c,\text{nk}} - \x_{\text{k}}}_2$ is assumed to be less than $\epsilon$. By substituting \Cref{eq:simplified} into \Cref{eq:upscale_uk}, we derive the approximation as
\begin{align}
    \ensmean{\Theta^{\left(i\right)}}_Y \left(\x\right) &= \left\langle T_\epsilon^{\left(i\right)}\right\rangle_{\mathcal{W}_{\epsilon}^{\left(i\right)}\left(\x\right)} = \frac{1}{\left\vert Y \right\vert} \int_{\mathcal{W}_{\epsilon,\text{k}}^{\left(i\right)}\left(\x\right)} T_\epsilon^{\left(i\right)} \left(
    \y\right)\dd\y \nonumber \\
    &+\beta \phi^{\left(i\right)} \left[ T_\epsilon^{\left(i\right)}\bigg\vert_{\y = \x_k} + \grad{T_\epsilon^{\left(i\right)}} \bigg\vert_{\y = \x_k}^T \left(\x_{c,\text{nk}} - \x_{\text{k}}\right) \right] + \frac{1}{\left\vert Y \right\vert} \int_{\mathcal{W}_{\epsilon,\text{nk}}^{\left(i\right)}\left(\x\right)}  \mathcal{O}\left(\left(\y - \x_{\text{k}} \right)^2\right) \ \dd\y.
\end{align}
When $\beta \rightarrow 0$, we recover \Cref{eq:upscale_kernel_averaging}, demonstrating the consistency of the derived approximation method.

In summary, the upscaled kernel is defined as 
\begin{align}
\begin{cases}
     \ensmean{\Theta^{\left(i\right)}}_Y (\x) = \ensmean{T^{\left(i\right)}}_Y (\x), \quad \x \in \mathcal{V}_{\text{up}} \subset \Omega_{\text{up}}^{\text{old}}, \\
     \ensmean{\Theta^{\left(i\right)}}_Y \left(\x\right) = \frac{1}{\left\vert Y \right\vert} \int_{\mathcal{W}_{\epsilon,\text{k}}^{\left(i\right)}\left(\x\right)} T_\epsilon^{\left(i\right)} \left(
    \y\right)\dd\y + \beta \phi^{\left(i\right)} \left[ T_\epsilon^{\left(i\right)}\bigg\vert_{\y = \x_k} + \grad{T_\epsilon^{\left(i\right)}} \bigg\vert_{\y = \x_k}^T \left(\x_{c,\text{nk}} - \x_{\text{k}}\right) \right], \quad \x \in \mathcal{V}_{\text{fine}\rightarrow \text{up}}, \\
    \Theta^{\left(i\right)}_\epsilon(\x) = T^{\left(i\right)}_\epsilon(\x), \quad \x \in \mathcal{V}_{\text{fine}} \subseteq \Omega_{\text{fine}}^{\text{old}}.
\end{cases}
\end{align}

\subsection{Adaptive hybridization scheme}\label{sec:summary}
\begin{figure}[H]
\centerline{
 {\includegraphics[width=.8\textwidth]{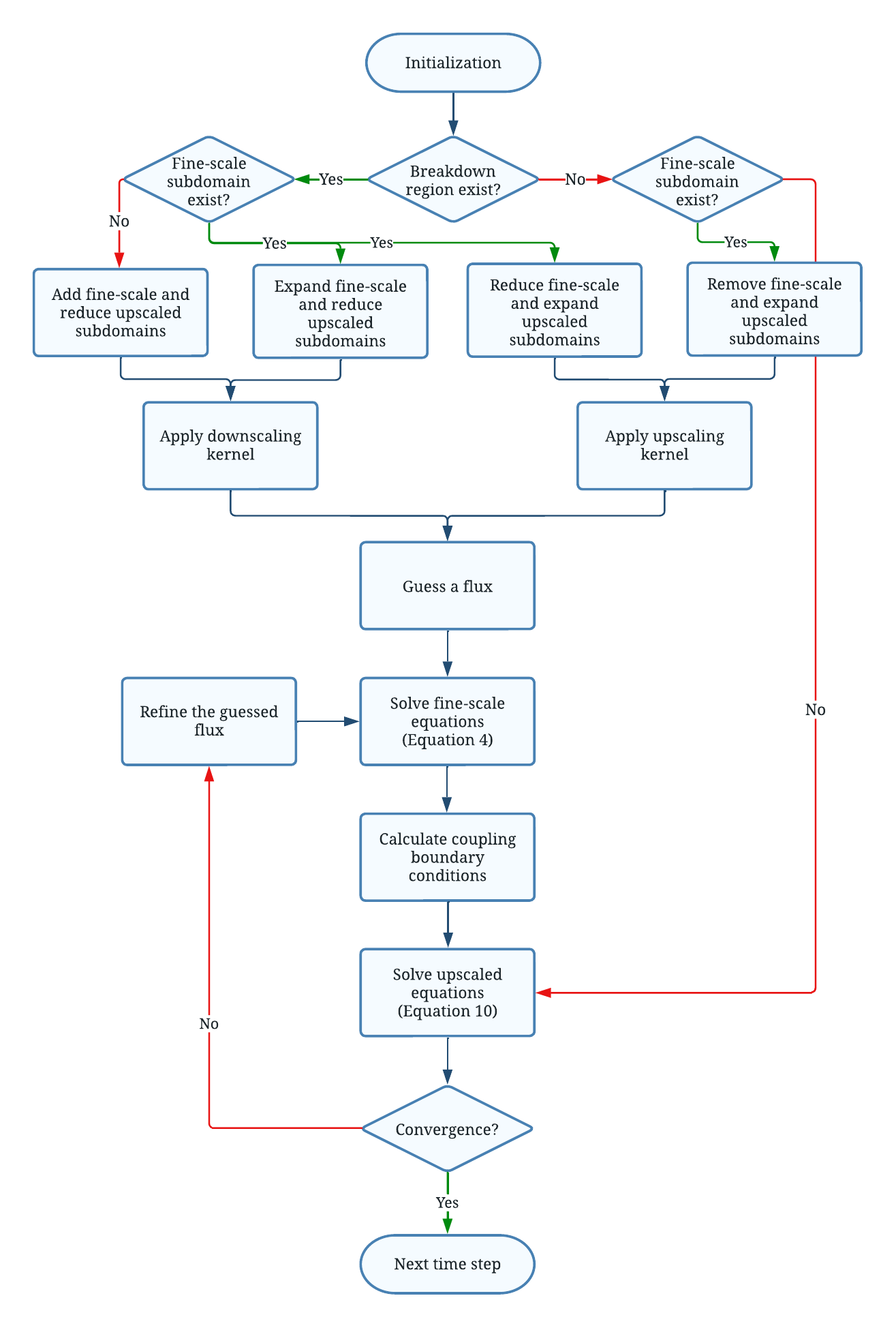}}}

\caption{Algorithm flow chart of the proposed automatic-dectecting and adaptive, two-sided hybrid method.}
\label{fig:flowchart}
\end{figure}

After defining the upscaling and downscaling kernels, the algorithm for adaptive hybridization is shown in \Cref{fig:flowchart}. Below, we discuss the main steps of the algorithm. 


\begin{enumerate}
    \item Initialize all parameters (e.g., dimensionless numbers, numerical parameters), as well as packing and cell temperatures.
    \item Check for the existence and boundaries of breakdown regions using \Cref{eq:AD_for_all_DNs} to define the boundaries of fine-scale and upscaled subdomains.
    \item Based on the existence of breakdown regions and the state of the subdomains, determine the appropriate next step:
    \begin{itemize}
        \item If a breakdown region exists and the boundaries of the subdomains remain unchanged, proceed to steps 4--9.
        \item If a breakdown region exists and the boundaries of the subdomains change, regenerate meshes for the respective subdomains with the appropriate resolution, apply the necessary kernels to modify the fine-scale and upscaled subdomains accordingly, and then proceed to steps 4--9.
        \item If a breakdown region does not exist but a fine-scale subdomain exists, regenerate meshes for the upscaled subdomain, apply the appropriate kernels to modify the fine-scale subdomains, and solve the upscaled equations (\Cref{eq:upscaled_goven_eqs}) directly.
        \item If neither a breakdown region nor a fine-scale subdomain exists, solve the upscaled equations (\Cref{eq:upscaled_goven_eqs}) directly.
    \end{itemize}
    \item Solve the fine-scale equations using an initial guess for the unresolved flux $q^{\left(p,n\right)}$.
    \item {Evaluate the average fine-scale flux of the packing materials using Taylor expansion} (\Cref{eq:taylor-hc-flux}).
    \item Solve the upscaled equations (\Cref{eq:upscaled_goven_eqs}) with the calculated average packing materials flux.
    \item Compute the average packing temperature using Taylor expansion (\Cref{eq:taylor-hc-temp}).
    \item {Calculate the error in the continuity of the average temperature, $\mathcal{F}$, at the coupling boundary using the following expression:}

    \begin{align}
    \mathcal{F} = \max \left(\ensmean{T^{\left(p\right)}}_{Y}(\mathbf{x}^{-}_l) - \langle T_\epsilon^{\left(p\right)} \rangle_{\mathcal{W}_{\epsilon}^{\left(p\right)}\left(\x^+_l\right)}, \ensmean{T^{\left(p\right)}}_{Y}(\mathbf{x}^{-}_r) - \langle T_\epsilon^{\left(p\right)} \rangle_{\mathcal{W}_{\epsilon}^{\left(p\right)}\left(\x^+_r\right)}\right)
    \end{align}
    \item {If $\max(\norm{{\mathcal{F}}}_\infty, \norm{{\mathcal{F}}}_2) > \epsilon_{tol}$, where $\epsilon_{tol}$ is a specific tolerance, update the unresolved flux $q^{\left(p,n\right)}$ by applying a root-finding algorithm (e.g., Broyden's method) and iterate through steps 4--9.}
\end{enumerate}

A detailed flow chart of the algorithm is provided in \Cref{fig:flowchart}.

\section{Numerical Verification}
\label{sec:res_discuss}
\subsection{Simulations Setup}
We test the accuracy of the developed automatic-detecting and adaptive hybrid formulation on four test cases modeling thermal runaway in a two-dimensional battery pack. To verify the coupling and the algorithm, we compare the results of fine-scale and hybrid simulations for all four cases to ensure that the error introduced by the hybridization strategy is bounded by the theoretical upscaling error, $\epsilon$ \cite{Pietrzyk2023-ou}. All test cases are run on the same computational domain with different initial conditions and/or parameter values  and involve a ``breakdown'' region that can evolve in space and/or time to test the algorithm on setups of increasing complexity (and detailed in the following sections). Pietrzyk \emph{et al.} \cite{Pietrzyk2023-ou} demonstrated that the upscaled equations are valid as long as the thermal runaway problem satisfies the applicability regimes of \Cref{eq:regimes}. The space-time locations in which such conditions are not met identify the ``breakdown'' region and are detected through criteria discussed in Section \ref{sec:ad-formulation}.

\begin{figure}[hbpt!]
\centerline{
 {\includegraphics[width=\textwidth]{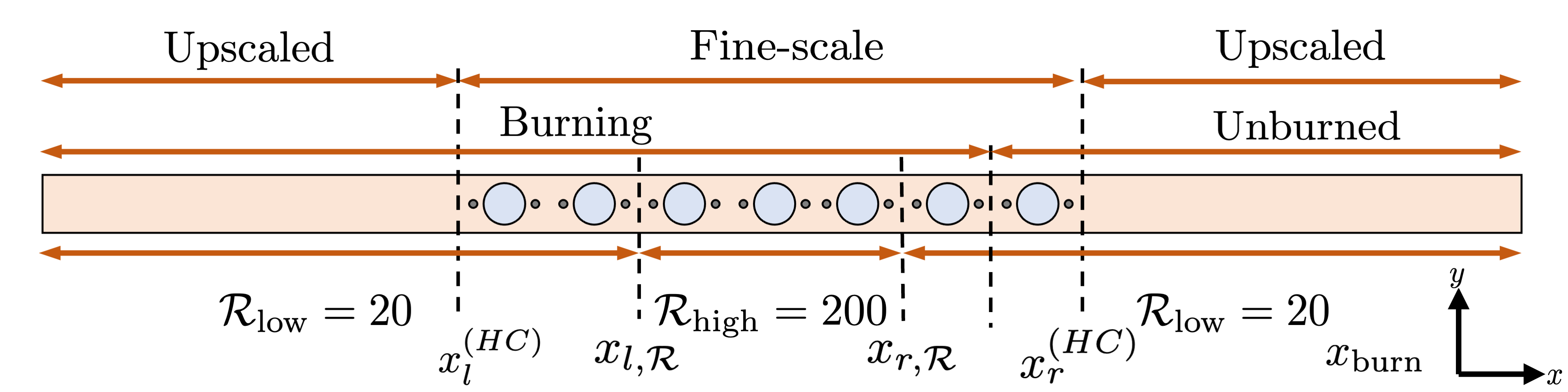}}}

\caption{{Schematic diagram illustrating the simulation setup.}}
\label{fig:setup-diag}
\end{figure}

We consider a two-dimensional computational domain of size $1\cross \epsilon$ with $20$ unit cells in the $x$-direction ($N_{x}^{\left(c\right)}=20$)  and one unit cell in the $y$-direction ($N_{y}^{\left(c\right)}=1$), as illustrated in \Cref{fig:setup-diag}, i.e. $\epsilon=0.05$.  Periodic boundary conditions and zero gradient boundary conditions are applied along boundaries in $x$- and $y$-direction, respectively. The parameters defining the unit-cell geometry in Figure \ref{fig:finescale-domain} follow those outlined in \cite{Yao2023-zi}, and are detailed in \Cref{tab:unit-cell-geom-params} for completeness.

\begin{table}[ht!]
\centering
{\caption{{Parameter used to define the unit cell in the simulations.}}
\begin{tabular}{l|ccccc}
\hline \hline
\multirow{ 2}{*}{Parameter} & $\hat{r}_\epsilon^{\left(c\right)}$ & $\hat{r}_\epsilon^{\left(w\right)}$  & $d^{\left(cc\right)}_\epsilon$  & $d^{\left(1\right)}_\epsilon$  & $d^{\left(2\right)}_\epsilon$   \\
 & [\si{L}] & [\si{L}] & [\si{L}] & [\si{L}] & [\si{L}]  \\
\hline
Value & 0.009 & 0.003 & 0.009 & 0.001 & 0.002 \\
\hline \hline
\end{tabular}}
\label{tab:unit-cell-geom-params}
\end{table}

The computational domain is spatially heterogeneous, i.e. it contains battery cells with two different values of the \textit{burn} power flux  $\hat{\Pi}_{\text{burn}}$ (see \Cref{eq:Scales}), which  reflects, e.g., different thermal behaviors because of potential manufacturing defects, chemistry, and aging. Specifically, we indicate with $\hat{\Pi}_{\text{burn},\text{low}}$ and $\hat{\Pi}_{\text{burn},\text{high}}$ such two values, where the latter is 10 times larger than the former, and their spatial distribution across the battery pack is defined as follows (see Figure \ref{fig:setup-diag}),
\begin{align}\label{eq:R-finescale}
    \hat{\Pi}_{\text{burn}} = 
    \begin{cases}
    \hat{\Pi}_{\text{burn},\text{low}}=\hat{\Pi}_{\text{burn},0}, \quad {x} < {x}_{l,\mathcal{R}} \quad \text{and} \quad {x} > {x}_{r,\mathcal{R}}, \\
    \hat{\Pi}_{\text{burn},\text{high}}=10\hat{\Pi}_{\text{burn},0}, \quad {x}_{l,\mathcal{R}} \le {x} \le {x}_{r,\mathcal{R}},
    \end{cases}
\end{align}
where $\hat{\Pi}_{\text{burn},0}$ is selected such that
\begin{align}\label{eq:Pi_burn0}
    &\hat{\Pi}_{\text{burn},0} := \ddfrac{\hat{T}_{max} \hat{k}^{\left(p\right)}}{\max{(\hat{\mathcal{L}}_{x}, \hat{\mathcal{L}}_{y})}\hat{\ell}}.
\end{align}
\noindent It is important to emphasize that Equation \eqref{eq:Pi_burn0} corresponds to selecting  $\mathcal{R}=\mathcal{R}_{\text{low}}:=\epsilon^{-1}$ (through \Cref{eq:dimless_groups}). This value of $\mathcal{R}$ falls within the applicability regimes (\Cref{eq:regimes}) derived by \cite{Pietrzyk2023-ou}. However, for those cells where $ \hat{\Pi}_{\text{burn}}= \hat{\Pi}_{\text{burn},\text{high}}$, $\mathcal{R}=10\epsilon^{-1}=10\mathcal{R}_{\text{low}}$, the applicability conditions of the upscaled model Equations \eqref{upscaled}-\eqref{eq:effectivecoefficients} are violated, a ``breakdown'' region exists, and hybridization is necessary. 
The fine-scale heterogeneity represented by Equation \eqref{eq:R-finescale} can be described at the continuum-scale by the following approximation of a step-function:
\begin{align}
\label{eq:R_cal}
\mathcal{R}(x, x_{\mathcal{R}}) = 
\begin{cases}
\left(\dfrac{\mathcal{R}_{\text{high}} + \mathcal{R}_{\text{low}} }{2}\right) - \left(\dfrac{\mathcal{R}_{\text{high}} - \mathcal{R}_{\text{low}} }{2}\right)\tanh\left(\zeta(x-x_{\mathcal{R}}) \right), \quad x < 0\\
\left(\dfrac{\mathcal{R}_{\text{high}} + \mathcal{R}_{\text{low}} }{2}\right) - \left(\dfrac{\mathcal{R}_{\text{high}} - \mathcal{R}_{\text{low}} }{2}\right)\tanh\left(\zeta(x+x_{\mathcal{R}}) \right), \quad x \ge 0
\end{cases}
\end{align}
where $\zeta=180$ is a constant used to approximate a step function, $x_{\mathcal{R}} = (x_{l,\mathcal{R}} + x_{r,\mathcal{R}})/2$ is the center of the region with high heat generation rate, and $\mathcal{R}_{\text{high}}= 10\mathcal{R}_{0}$ and $\mathcal{R}_{\text{low}} = \mathcal{R}_{0}$ represent regions with high and low heat generation rates, respectively.

The four test cases are designed such that $\mathcal{R}$ is a known function of space and time, and the breakdown region can both expand and contract. Before proceeding to a detailed discussion of the four cases, additional information (including initial conditions) about the problem setup are provided below.

\textbf{Initial Conditions: Pore-scale.} To model battery thermal runaway at the pore-scale, we explicitly define the burning and unburned regions of the battery cells at $t=0$, by defining the spatially heterogenous $\Pi\left(T_{\epsilon}^{\left(c\right)}, {x}\right)$ as 
\begin{align}\label{eq:pore-scaleIC}
\Pi\left(T_{\epsilon}^{\left(c\right)}, {x}\right) &=
\begin{cases}
\Pi_{\text{NB}}\left(T_{\epsilon}^{\left(c\right)}\right), \text{ for } {x} > {x}_{\text{burn}}, \\
\Pi_{\text{FB}}\left(T_{\epsilon}^{\left(c\right)}\right), \text{ for } {x} \le {x}_{\text{burn}},
\end{cases}
\end{align}
where $x_{\text{burn}}$ separates the burning and unburned regions, $\Pi_{\text{NB}}\left(T_{\epsilon}^{\left(c\right)}\right)$ and $\Pi_{\text{FB}}\left(T_{\epsilon}^{\left(c\right)}\right)$ are the dimensionless fine-scale power flux source terms for unburned and burning battery cell defined as
\begin{subequations}
\begin{align}
\Pi_{\text{NB}}\left(T_{\epsilon}^{\left(c\right)}\right) &= \Pi_{\text{base}} + \frac{1}{2}\left\{\text{Erf}\left[A_1 T_{\epsilon}^{\left(c\right)} + B_1\right] + 1\right\}\left(1 - \Pi_{\text{base}}\right) \nonumber \\ 
& - \frac{1}{2}\left\{\text{Erf}\left[A_2 T_{\epsilon}^{\left(c\right)} + B_2\right] + 1\right\}\\
\Pi_{\text{FB}}\left(T_{\epsilon}^{\left(c\right)}\right) &= \Pi_{\text{burn}} - \frac{1}{2}\left\{\text{Erf}\left[A_2 T_{\epsilon}^{\left(c\right)} + B_2\right] + 1\right\}.\label{eq:new_pi_term_dimless_burn}
\end{align}
\end{subequations}
where $\Pi_{\text{FB}}$ can be obtained  from \Cref{eq:new_pi_term_dimless} by setting $\Pi_{\text{base}} = \Pi_{\text{burn}} = 1$. 
It is worth emphasizing that this setup allows for heat generated from the burned cells to propagate to the unburned cells, creating a thermal front causing originally unaffected battery cells to undergo thermal runaway as the thermal front reaches them.

\textbf{Initial Conditions: Continuum-scale.} The upscaled power flux source terms at $t=0$ corresponding to the initial condition \eqref{eq:pore-scaleIC} is
\begin{align}
\overline{\Pi} \left(\langle T^{\left(c\right)} \rangle_{Y}, {x}\right) &= \ddfrac{ \overline{\Pi}_{\text{FB}} \left(\langle T^{\left(c\right)} \rangle_{Y}, {x}\right) -  \overline{\Pi}_{\text{NB}} \left(\langle T^{\left(c\right)} \rangle_{Y}, {x}\right)}{1+\exp\left(\gamma({x} - 
{x}_{\text{burn}})\right)} + \overline{\Pi}_{\text{NB}} \left(\langle T^{\left(c\right)} \rangle_{Y}, {x}\right), \label{eq:upscale_force_pi_term}
\end{align}
where $\gamma=180$ is an arbitrary constant to approximate a stepwise function at the continuum scale. $\overline{\Pi}_{\text{NB}} \left(\langle T^{\left(c\right)} \rangle_{Y}, {x}\right)$ and $\overline{\Pi}_{\text{FB}} \left(\langle T^{\left(c\right)} \rangle_{Y}, {x}\right)$ are the dimensionless upscaled power flux source term for unburned and burning battery cells, defined as
\begin{subequations}
\begin{align}
\overline{\Pi}_{\text{NB}} \left(\langle T^{\left(c\right)} \rangle_{Y}, {x}\right) &= \Pi_{\text{base}} + \frac{1}{2}\left\{\text{Erf}\left[\frac{A_1}{\phi^{\left(c\right)}} \langle T^{\left(c\right)} \rangle_{Y} + B_1\right] + 1\right\}\left(1 - \Pi_{\text{base}}\right)  \nonumber \\
&- \frac{1}{2}\left\{\text{Erf}\left[\frac{A_2}{\phi^{\left(c\right)}} \langle T^{\left(c\right)} \rangle_{Y} + B_2\right] + 1\right\}, \\
\overline{\Pi}_{\text{FB}} \left(\langle T^{\left(c\right)} \rangle_{Y}, {x}\right) &= \Pi_{\text{burn}} - \frac{1}{2}\left\{\text{Erf}\left[\frac{A_2}{\phi^{\left(c\right)}} \langle T^{\left(c\right)} \rangle_{Y} + B_2\right] + 1\right\}. 
\end{align}
\end{subequations}
Equation \eqref{eq:upscale_force_pi_term} allows to consistently define the transition zone between burned and unburned regions whenever such zone falls in the continuum domain.

{For all simulations, we specify an arbitrary end time of $t_{f}/\hat{t} = 0.2$, equivalent to $6350 \Delta t$. The reference timescale for heat transfer is given by
\begin{align}
    \hat{t} &= \ddfrac{\hat{\rho}^{\left(p\right)}\hat{C}^{\left(p\right)}\left(\max{(\hat{\mathcal{L}}_{x}, \hat{\mathcal{L}}_{y})}\right)^2}{\hat{k}^{\left(p\right)}},
\end{align}
where $\hat{t}$ is the reference time based on the properties of the packing maiterials. \Cref{tab:ref-val-scaling} summarizes the reference parameters for defining the dimensionless numbers in~\Cref{eq:dimless_groups}~\cite{Yao2023-zi}. The simulations are initialized with zeros for all fields.}

In the following, we test the accuracy of the two-dimensional fixed hybrid coupling (Section~\ref{sec:acc-ts-fixed}),  the detection criterion (Section~\ref{sec:acc-ad}), the downscaling kernel for hybrids with expanding fine-scale domain (Section~\ref{sec:acc-adaptive-expansion}), and the upscaling kernel for hybrids with expanding continuum-scale domain (Section~\ref{sec:acc-adaptive-reduction}) in Test Cases 1, 2, 3 and 4, respectively. The parameters of each case are summarized in Table~\ref{tab:val-sim-params}.

\begin{table}[ht!]
\centering
{\caption{{Reference parameters for dimensionless numbers and nondimensionalizing the governing equations.}}
\begin{tabular}{l|cccccccc}
\hline \hline
\multirow{ 2}{*}{Parameter} & $\hat{\rho}^{\left(c\right)}$ & $\hat{\rho}^{\left(p\right)}$ & $\hat{C}^{\left(c\right)}$ & $\hat{C}^{\left(p\right)}$ & $\hat{k}^{\left(c\right)}$ & $\hat{k}^{\left(p\right)}$ & $\hat{T}_\infty$  \\
 & [\unit{ML\tothe{-3}}] & [\si{ML\tothe{-3}}] & [\unit{L\tothe{2}T\tothe{-2}\Theta\tothe{-1}}] & [\unit{L\tothe{2}T\tothe{-2}\Theta\tothe{-1}}] & [\unit{MLT\tothe{-3}\Theta\tothe{-1}}] & [\unit{MLT\tothe{-3}\Theta\tothe{-1}}] & [\unit{\Theta}] \\
\hline
Value & 2500 & 1500 & 900 & 1500 & 3 & 3 & 293 \\
\hline \hline
\multirow{ 2}{*}{Parameter} & $\hat{T}_{ref}$ & $\hat{T}_a$ & $\hat{T}_b$ & $\hat{T}_{s1}$ & $\hat{T}_{s2}$ & $\epsilon_{s1}$ & $\epsilon_{s2}$ \\
 & [\unit{\Theta}] & [\unit{\Theta}] & [\unit{\Theta}] & [\unit{\Theta}] & [\unit{\Theta}] & [-] & [-] \\
\hline
Value & 293 & 0 & 0 & 120 & 120 & 0.0005 & 0.0005 \\
\hline \hline
\end{tabular}}
\label{tab:ref-val-scaling}
\end{table}

\subsection{Case 1: Accuracy of two-sided fixed hybrid coupling}
\label{sec:acc-ts-fixed}

Prior to implementing formulations with automatic detection and adaptive capabilities, we first conduct  hybrid simulations with fixed coupling boundaries located at $x^{(HC)}_{l}$ and $x^{(HC)}_{r}$ to verify the accuracy of the two-sided coupling formulation. The errors between the fine-scale and hybrid simulations are expected to be bounded by the upscaling error, i.e. $\mathcal{O}(\epsilon) \sim 0.05$ \cite{Pietrzyk2023-ou}.

\begin{landscape}
\begin{figure}[H]
    \centering
    {\includegraphics[width=1.5\textwidth,trim=0 7.3cm 0 0, clip]{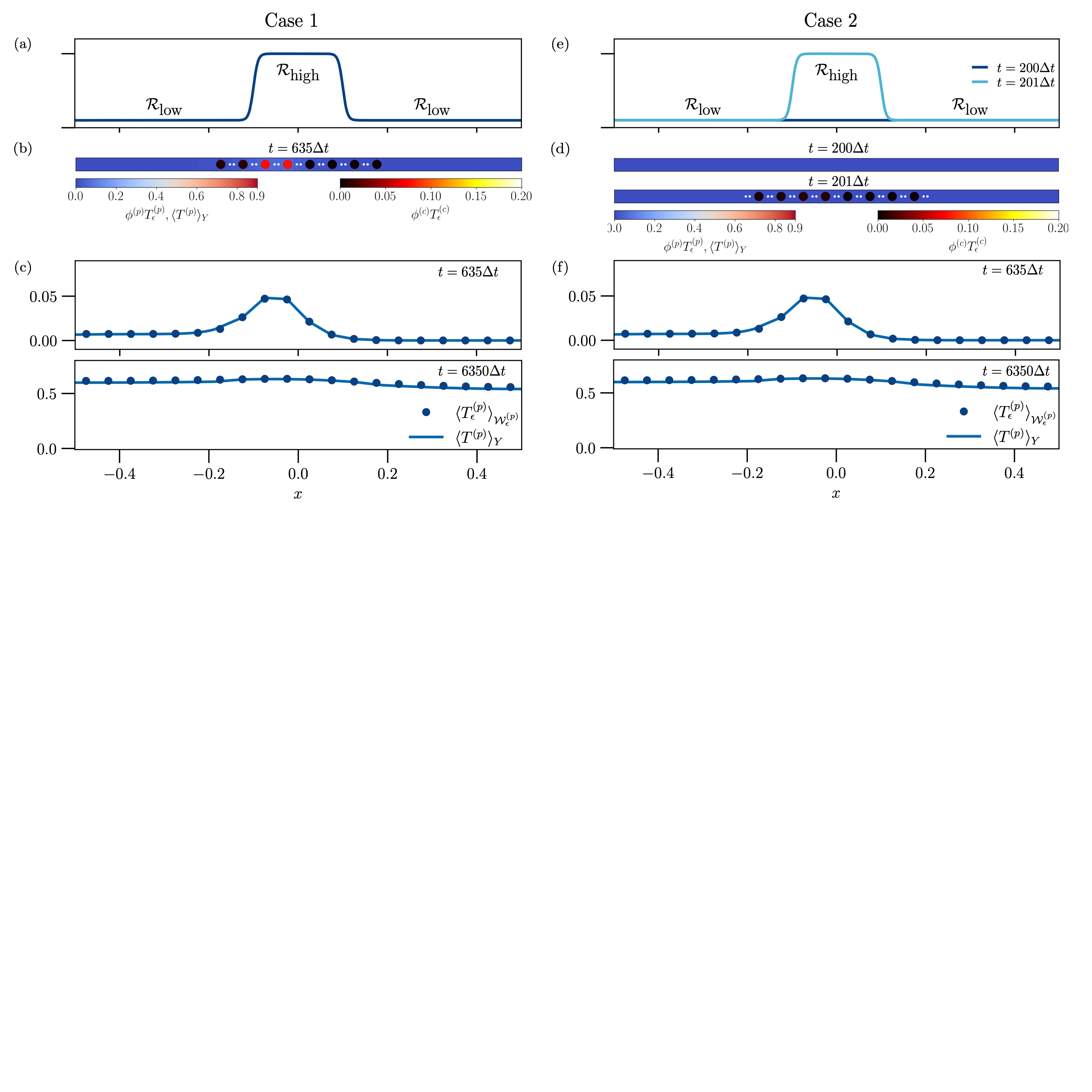}}
     \caption{(a) and (e) Dimensionless number $\mathcal{R}$ plotted as a function of $x$ for Cases 1 and 2, respectively; (b) and (d) simulation snapshots for Cases 1 and 2. (c) and (f) The average packing temperature along the centerline ($y=0$) is computed at $t= 635\Delta t$ and $t=6350 \Delta t$ for Cases 1 and 2, respectively.}
     \label{fig:case1_2}
\end{figure}
\end{landscape}

\Cref{tab:val-sim-params} summarizes the parameters used to set up the simulations. Initially, we define a breakdown region between $x_{l,\mathcal{R}}$ and $x_{r,\mathcal{R}}$ characterized by a high rate of heat generation. Following Yao \emph{et al.}~\cite{Yao2023-zi}, we set the coupling boundaries 1.5$\epsilon$ away from $x_{l,\mathcal{R}}$ and $x_{r,\mathcal{R}}$ to ensure the accuracy of the two-sided hybrid coupling formulation. \Cref{fig:case1_2}(a) illustrates $\mathcal{R}$ as a function of $x$, separating the breakdown and non-breakdown regions.

The simulation snapshot (\Cref{fig:case1_2}(b)) demonstrates the presence of two coupling boundaries that encompass the entire breakdown region. \Cref{fig:case1_2}(c) shows the average centerline packing temperature from both fine-scale and hybrid simulations. At $t=635 \Delta t$ and $6350 \Delta t$, the discrepancies between the fine-scale and hybrid simulations are negligible, as anticipated.

To quantify the accuracy of the coupling, the errors at time $t$ are computed as 
\begin{align}
\label{eq:err-cal}
    &err (\x, t) = \abs{\ensmean{T^{\left(i\right)}_\epsilon}_{\mathcal{W}_{\epsilon}^{\left(i\right)}\left(\x\right)}(t) - T^{\left(i\right)}_{HC} (\x, t)},
\end{align}
where $i=c$ or $p$ represent the cell or packing temperature, respectively, $\ensmean{T^{i}_\epsilon}_{\mathcal{W}_{\epsilon}^{\left(i\right)}\left(\x\right)}(t)$ is the averaged fine-scale cell or packing temperatures at location $\x$ and time $t$, and $T^{\left(i\right)}_{HC}$ is the average temperature in the hybrid simulations such that
\begin{align}
    T^{\left(i\right)}_{HC}(\x, t) = 
    \begin{cases}
        \ensmean{T^{\left(i\right)}_\epsilon}_{\mathcal{W}_{\epsilon}^{\left(i\right)}\left(\x\right)}(t), \quad \x \in \Omega_{\text{fine}}, \\
        \ensmean{T^{\left(i\right)}}_Y(\x, t), \quad \x \in \Omega_{\text{up}}.
    \end{cases}
\end{align}
To show that the errors are consistently smaller than the upscaling error for the entire simulations, we compute the maximum spatial errors as a function of time such that 
\begin{align}
    err_{x} (t) = \max_{\x} \left(err (\x, t) \right), \quad 0\le t \le t_{f},
\end{align}
where $t_{f} = 6350 \Delta t$. \Cref{fig:max_err} shows the maximum spatial errors as a function of time. For Case 1, the packing and cell temperature errors are constantly smaller than the upscaling error (indicated by the black dashed line), which confirms the accuracy of the two-sided hydrid coupling. 

\begin{table}[ht!]
\centering
\caption{{Simulation parameters used in the test cases to validate the proposed hybrid algorithm. $h_{up,min}$ and $h_{fine,min}$ are the minimum grid resolution for upscaled and fine-scale simulations, respectively.}}
\begin{tabular}{c|c|c|c|c}
\hline \hline
 & Case 1 & Case 2  & Case 3& Case 4 \\
 & (\Cref{sec:acc-ts-fixed}) & (Section~\ref{sec:acc-ad}) & (\Cref{sec:acc-adaptive-expansion})& (Section~\ref{sec:acc-adaptive-reduction})\\
\hline
{$\Delta{t}/\hat{t}$ ($10^{-5}$)}  &  \multicolumn{4}{c}{$3.15$} \\
{$N_x \times N_y$} & \multicolumn{4}{c}{$20 \times 1$}  \\
{($h_{up,min}$, $h_{fine,min}$)}  & \multicolumn{4}{c}{($1.00 \times 10^{-2}$, $2.50\times10^{-4}$)}  \\
{No. of $\Delta{t}$}   & \multicolumn{4}{c}{$6350$}   \\
{ $x_{\text{burn}}$ } & \multicolumn{4}{c}{0.0} \\\hline
{($x_{l}^{(HC)}$, $x_{r}^{(HC)}$) at $t=0$} & (-0.1875, 0.1875) & N.A & (-0.1375, 0.1375) & (-0.4500, 0.4500)\\
{($x_{l}^{(HC)}$, $x_{r}^{(HC)}$) at $t = t_f$} & (-0.1875, 0.1875) & (-0.2125, 0.2125) & (-0.4125, 0.4125) & (-0.1625, 0.1625)\\
{($x_{l,\mathcal{R}}$, $x_{r,\mathcal{R}}$) at $t = 0$}  & (-0.10, 0.10) & N.A & (-0.05, 0.05) & (-0.30, 0.30) \\
{($x_{l,\mathcal{R}}$, $x_{r,\mathcal{R}}$) at $t = t_f$} & (-0.10, 0.10) & (-0.10, 0.10) & (-0.30, 0.30) & (-0.10, 0.10) \\
\hline \hline
\end{tabular}%
\label{tab:val-sim-params} 
\end{table}

\subsection{Case 2: Accuracy of automatic-detecting fixed hybrid coupling}
\label{sec:acc-ad}

As previously discussed, the definition of an automatic detection strategy that continuously monitors whether any dimensionless numbers exceed their defined applicability regimes is critical for automated hybridization algorithms (\Cref{sec:ad-formulation}). The formulation of automatic detection is evaluated based on two critical aspects: (1) reliable detection and (2) accurate mapping of information between fine-scale and upscaled subdomains. Reliable detection not only refers to the ability to detect changes but also involves generating accurate boundaries for the fine-scale subdomain. These boundaries should be optimally sized to both avoid excessive computational costs associated with overly large subdomains and prevent inaccuracies in hybrid simulations due to overly small subdomains. 
This section  focuses on evaluating the capability of auto-detection to accurately generate and maintain an appropriate fine-scale subdomain within an upscaled simulation.

The parameters used to simulate Case 2 are summarized in \Cref{tab:val-sim-params}. Initially, the simulation starts with  the upscaled subdomain occupying the full computational domain. 

A hybrid simulation was triggered by imposing a region of $\mathcal{R}_{\text{high}}$ around $x_{\mathcal{R}} = 0$ as defined by \Cref{eq:R_cal}. \Cref{fig:case1_2}(d) shows the distribution of $\mathcal{R}$ before and after the change in $\mathcal{R}$. The simulation is initialized with $\mathcal{R} = 20$ for all values of $x$. When the magnitude of $\mathcal{R}$ remains within the applicability regime for $t \le 200\Delta{t}$, only the upscaled equations were solved. At $t=201\Delta{t}$, a region of $\mathcal{R}_{\text{high}}=200$ was introduced for $-0.1 \le x \le 0.1$, resulting in a violation of the applicability regime for $\mathcal{R}$ and leading to the introduction of a fine-scale subdomain. As such, the simulation transitioned to a hybrid simulation as shown in the snapshots at $t = 200 \Delta t$ and $t = 201 \Delta t$ (\Cref{fig:case1_2}(e)).  Visual inspection of the distribution of $\mathcal{R}$ indicates that the left and right boundaries ($x_{l}^{\left(HC\right)}$ and $x_{r}^{\left(HC\right)}$) are approximately located at $x=-0.15$ and $x=0.15$, where $\mathcal{R} \approx 20$, respectively. To ensure adequate separation between the coupling and breakdown boundaries,  a buffer distance $x_{dist}$  greater than or equal to $1.5\epsilon$ was maintained, resulting in the left and right coupling boundaries being located approximately at $x=-0.225$ and $x=0.225$. According to \Cref{eq:AD_for_all_DNs}, the left and right coupling boundaries are calculated as $x=-0.2125$ and $x=0.2125$, demonstrating a close agreement with the visually determined boundary locations. 


\Cref{fig:case1_2}(f) shows the centerline of the average packing temperature  at two different instances in time $t=635 \Delta t$ and $6350 \Delta t$ after the introduction of a fine-scale subdomain. The negligible differences between fine-scale and hybrid simulations demonstrate the accuracy of the automatic detection developed. Similarly, as discussed in~\Cref{sec:acc-ts-fixed}, \Cref{fig:max_err} shows the maximum spatial errors over time. The maximum errors in the packing and cell temperatures consistently remain below the upscaling error threshold.

\subsection{Case 3: Accuracy of hybrid simulation with expanding fine-scale subdomain}
\label{sec:acc-adaptive-expansion}

The size of the breakdown region often varies over time. For instance, the heat generation rates in burning battery cells may increase, leading to varying $\mathcal{R}$ values. With fixed coupling boundaries, simulations must accommodate the largest potential breakdown region, even if this scenario is transient or short-term. As a result, this approach can lead to unnecessary computational costs, thereby reducing the efficiency and benefit of hybrid simulations. To minimize computational costs while maintaining accuracy, the size of the fine-scale subdomain must be adjusted dynamically to match the changing sizes of the breakdown region. This section focuses on demonstrating the capabilities of the developed adaptive hybrid formulation to expand the fine-scale subdomain. The accuracy of the adaptive hybrid scheme is assessed  against the corresponding fine-scale simulations.

\begin{landscape}
\begin{figure}[H]
    \centering
      {\includegraphics[width=1.5\textwidth,trim=0 7.3cm 0 0, clip]{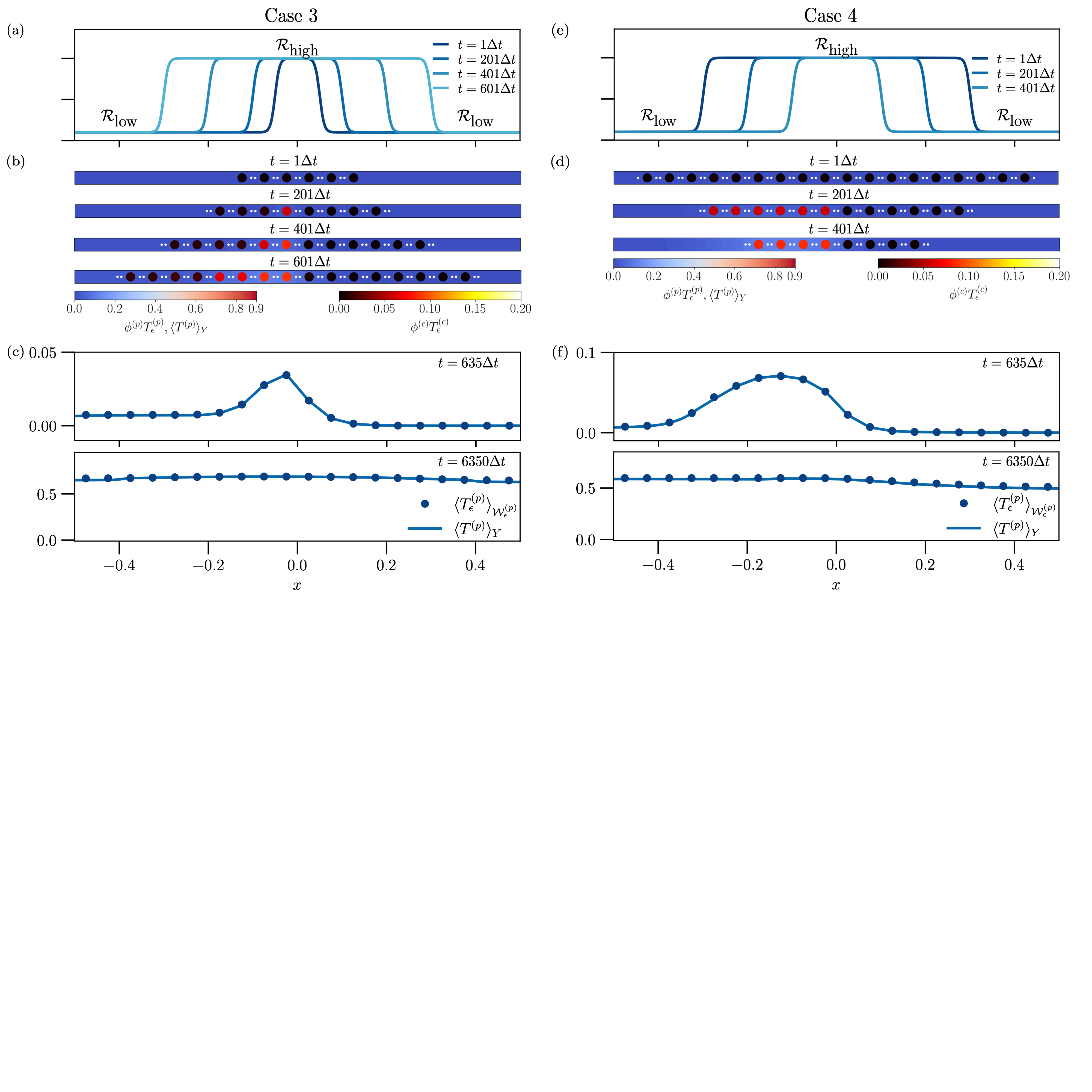}}
     \caption{(a) and (e): Dimensionless number $\mathcal{R}$ plotted as a function of $x$ in Cases 3 and 4, respectively. (b) and (d): simulation snapshots for Cases 3 and 4. (c) and (f): The average packing temperature along the centerline ($y=0$) is calculated at  $t= 635\Delta t$ and $t=6350 \Delta t$ for Cases 3 and 4.}
     \label{fig:case3_4}
\end{figure}
\end{landscape}

To model the varying size of the breakdown region, we define a function that adjusts the dimensionless number $\mathcal{R}$ over time. We modify the centers of the breakdown region $\x_{\mathcal{R}}$ such that 
\begin{align}
    x_{\mathcal{R}} = 
    \begin{cases}
    0.05 &  t \le 200\Delta{t}, \\
    0.10 & 200 \Delta{t} < t \le 400\Delta{t}, \\
    0.20 & 400 \Delta{t} < t \le 600\Delta{t}, \\
    0.30 & t > 600 \Delta{t}.
    \end{cases}
\end{align}
As $\mathcal{R}$ varies over time, the size of the fine-scale subdomain is dynamically expanded to accurately reflect the change in $\mathcal{R}$ and size of the breakdown region. \Cref{fig:case3_4}(a) illustrates the distribution of $\mathcal{R}$ as a function of $x$ for different values of $x_{\mathcal{R}}$, demonstrating how the breakdown region evolves over time.

\Cref{fig:case3_4}(b) illustrate the evolution of the fine-scale subdomain size during the hybrid simulations to accurately reflect the changes in $\mathcal{R}$. To evaluate the accuracy of the simulations, \Cref{fig:case3_4}(c) shows the centerline of the average packing temperature at $t=635 \Delta t$ and $6350 \Delta t$. The differences between fine-scale and hybrid simulations are almost negligible, as indicated by the overlapping lines. Similarly to previous sections, we also examine the temporal evolution of the maximum spatial errors of the average packing and cell temperatures (\Cref{fig:max_err}). Despite numerous expansions of the fine-scale subdomains throughout the hybrid simulations, the maximum spatial errors remain well within the upscaling errors, demonstrating the robustness and accuracy of the hybrid simulations. 

\subsection{Case 4: Accuracy of hybrid simulation with contracting fine-scale subdomain}
\label{sec:acc-adaptive-reduction}

\Cref{sec:acc-adaptive-expansion} demonstrated cases involving the expansion of fine-scale subdomains. Conversely, the size of the breakdown region may also decrease over time. For instance, the heat generation rates of burning battery cells tend to diminish after reaching their peak. Consequently, an adaptive hybrid simulation can not only expand but also contract the fine-scale subdomain as needed. This section will focus on scenarios that involve reducing the fine-scale subdomain and expanding the upscaled subdomains, utilizing the upscaled kernel developed in \Cref{sec:upscale_ker}.

\begin{figure}[H]
\centerline{
 {\includegraphics[width=0.7\textwidth]{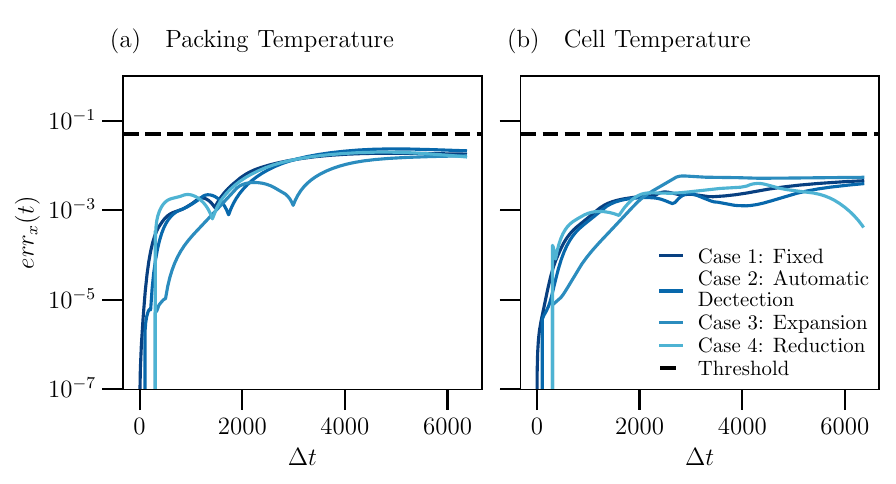}}}

\caption{Maximum spatial errors as a function of time for all test cases: (a) packing, and (c) cell temperatures.}
\label{fig:max_err}
\end{figure}

To mimic a reduced breakdown region, we initialize the simulation with a wider breakdown region, as shown in \Cref{fig:case3_4}(d). As time evolves, we deliberately reduce the size of the breakdown region by varying $x_{\mathcal{R}}$ such that
\begin{align}
    x_{\mathcal{R}} = 
    \begin{cases}
    0.30 &  t \le 200\Delta{t}, \\
    0.20 & 200\Delta{t} < t \le 400\Delta{t}, \\
    0.10 & t > 400\Delta{t}.
    \end{cases}
\end{align}
\Cref{fig:case3_4}(e) illustrates the time evolution of the fine-scale subdomains as $\x_{\mathcal{R}}$ changes. As the breakdown region diminishes, the size of the fine-scale subdomain decreases, demonstrating the capability to modify subdomain sizes appropriately. \Cref{fig:case3_4}(f) shows the centerline of the average packing temperature for both fine-scale and hybrid simulations at $t=635 \Delta t$ and $6350 \Delta t$. These lines were nearly perfectly aligned with each other, indicating that the differences between the average packing temperatures of fine-scale and hybrid simulations are minimal. \Cref{fig:max_err} shows the maximum spatial error as a function of time, clearly showing that the maximum spatial errors consistently remain within the bounds of the upscaling error.

\section{Conclusion}
We have developed an automatic-detecting and adaptive, non-intrusive two-sided hybrid method for multiscale heat transfer to couple fine-scale and upscaled models while ensuring that the coupling error is bounded by the theoretical upscaling error. This method includes strategies to constantly verify whether dimen-
sionless numbers remain within the applicability regimes of the continuum model. In practical applications, this adaptive hybrid method is most computationally efficient when the fine-scale subdomain constitutes a relatively small fraction of the overall domain, as larger fine-scale regions reduce computational savings. To enable adaptive hybrid simulations, we developed two kernels: an upscaling kernel for reducing the fine-scale subdomain and a downscaling kernel for expanding it. These kernels enable dynamic mapping of values between the fine-scale and upscaled subdomains. The hybrid method was numerically implemented using \fenics. To assess the accuracy of our hybrid method, we simulated a thermal runaway scenario within a battery pack, characterized by a spatially heterogenous distribution of heat generation rate, such that the applicability conditions of the upscaled model describing heat transport in the packing and cell domains were violated and a breakdown region could be identified. We conducted four different test cases to validate various aspects of the hybrid method: Case 1 tested the extension of the two-sided fixed hybrid method from the one-sided approach; Case 2 examined the accuracy of the auto-detecting feature, Case 3 evaluated the capability of the downscaled kernel to expand the fine-scale subdomain; and Case 4 assessed the effectiveness of the upscaled kernel in reducing the fine-scale subdomain. For all scenarios, comparisons of average temperatures at $t=635\Delta t$ and $t=6350 \Delta t$ showed that the differences between fine-scale and hybrid simulations were negligible. Additionally, maximum spatial errors were monitored over time, consistently staying below the upscaling errors.

\section*{Acknowledgments}
This material is based upon work supported by the Defense Advanced Research Projects Agency (DARPA) under Agreement No. HR00112090061. The views, opinions and/or findings expressed are those of the authors and should not be interpreted as representing the official views or policies of the Department of Defense or U.S. Government.

\appendix
\section{The Closure Problems}
\label{subsection:Appendix_E_Closure_Problems}
We present the formulation of the closure problems which are necessary for solving the upscaled equations. Identical formulation can be found in Pietrzyk \emph{et al.} \cite{Pietrzyk2023-ou} and Yao \emph {et al.} \cite{Yao2023-zi}. For these closure problems,  periodic boundary conditions are imposed at the edges of the unit cell such that $\langle \chi^{\left(p\right)\left[i\right]} \rangle_{\mathcal{B}^{\left(p\right)}} = 0$, $\langle \chiv^{\left(p\right)\left[3\right]} \rangle_{\mathcal{B}^{\left(p\right)}} = \bm{0}$, $\langle \chi^{\left(c\right)\left[1\right]} \rangle_{\mathcal{B}^{\left(c\right)}} = 0$, and $\langle \chiv^{\left(c\right)\left[2\right]} \rangle_{\mathcal{B}^{\left(c\right)}} = \bm{0}$ for $i \in \{1,2\}$.

The closure problem for $\chi^{\left(p\right)\left[1\right]}$ is given by
\begin{subequations}
\neweq{}{-\frac{\mathcal{Q}}{|\mathcal{B}^{\left(p\right)}|}|\Gamma^{\left(pw\right)}| - k^{\left(p\right)} \nabla_{\xiv} \bm{\cdot} \nabla_{\xiv} \chi^{\left(p\right)\left[1\right]} = 0 \quad \text{for } \xiv \in \mathcal{B}^{\left(p\right)},}
\neweq{}{-k^{\left(p\right)}\n^{\left(p\right)} \bm{\cdot} \nabla_{\xiv}\chi^{\left(p\right)\left[1\right]} = 0 \quad \text{for } \xiv \in \Gamma^{\left(pc\right)},}
\neweq{}{-k^{\left(p\right)} \n^{\left(p\right)} \bm{\cdot} \nabla_{\xiv}\chi^{\left(p\right)\left[1\right]} = \mathcal{Q} \quad \text{for } \xiv \in \Gamma^{\left(pw\right)}.}
\end{subequations}

The closure problem for $\chi^{\left(p\right)\left[2\right]}$ is given by

\begin{subequations}
\neweq{}{-\frac{\text{Bi}^{\left(p\right)}}{|\mathcal{B}^{\left(p\right)}|}|\Gamma^{\left(pc\right)}| - k^{\left(p\right)} \nabla_{\xiv} \bm{\cdot} \nabla_{\xiv} \chi^{\left(p\right)\left[2\right]} = 0 \quad \text{for } \xiv \in \mathcal{B}^{\left(p\right)},}
\neweq{}{-k^{\left(p\right)}\n^{\left(p\right)} \bm{\cdot} \nabla_{\xiv}\chi^{\left(p\right)\left[2\right]} = \text{Bi}^{\left(p\right)} \quad \text{for } \xiv \in \Gamma^{\left(pc\right)},}
\neweq{}{-k^{\left(p\right)}\n^{\left(p\right)} \bm{\cdot} \nabla_{\xiv}\chi^{\left(p\right)\left[2\right]} = 0 \quad \text{for } \xiv \in \Gamma^{\left(pw\right)}.}
\end{subequations}

The closure problem for $\chiv^{\left(p\right)\left[3\right]}$ is given by

\begin{subequations}
\neweq{}{-k^{\left(p\right)} \nabla_{\xiv} \bm{\cdot} \left(\I + \nabla_{\xiv} \chiv^{\left(p\right)\left[3\right]}\right) = \mathbf{0} \quad \text{for } \xiv \in \mathcal{B}^{\left(p\right)},}
\neweq{}{-k^{\left(p\right)} \n^{\left(p\right)} \bm{\cdot} \left(\I + \nabla_{\xiv}\chiv^{\left(p\right)\left[3\right]}\right) = \mathbf{0} \quad \text{for } \xiv \in \Gamma^{\left(pc\right)},}
\neweq{}{-k^{\left(p\right)} \n^{\left(p\right)} \bm{\cdot} \left(\I + \nabla_{\xiv}\chiv^{\left(p\right)\left[3\right]}\right) = \mathbf{0} \quad \text{for } \xiv \in \Gamma^{\left(pw\right)}.}
\end{subequations}

The closure problem for $\chi^{\left(c\right)\left[1\right]}$ is given by
\begin{subequations}
\neweq{cell_closure0_1}{\frac{\text{Bi}^{\left(c\right)} \varrho \varsigma}{|\mathcal{B}^{\left(c\right)}|}|\Gamma^{\left(pc\right)}| - k^{\left(c\right)} \varrho \varsigma \nabla_{\xiv} \bm{\cdot} \nabla_{\xiv} \chi^{\left(c\right)\left[1\right]} = 0 \quad \text{for } \xiv \in \mathcal{B}^{\left(c\right)},}
\neweq{cell_closure0_2}{-k^{\left(c\right)}\n^{\left(c\right)} \bm{\cdot} \nabla_{\xiv}\chi^{\left(c\right)\left[1\right]} = -\text{Bi}^{\left(c\right)} \quad \text{for } \xiv \in \Gamma^{\left(pc\right)}.}
\end{subequations}

The closure problem for $\chiv^{\left(c\right)\left[2\right]}$ is given by 
\begin{subequations}
\neweq{cell_closure1_1}{-k^{\left(c\right)} \varrho  \varsigma \nabla_{\xiv} \bm{\cdot} \left(\I + \nabla_{\xiv} \chiv^{\left(c\right)\left[2\right]}\right) = \mathbf{0} \quad \text{for } \xiv \in \mathcal{B}^{\left(c\right)},}
\neweq{cell_closure1_2}{-k^{\left(c\right)} \n^{\left(c\right)} \bm{\cdot} \left(\I + \nabla_{\xiv}\chiv^{\left(c\right)\left[2\right]}\right) = \mathbf{0} \quad \text{for } \xiv \in \Gamma^{\left(pc\right)}.}
\end{subequations}

\end{document}